\documentclass[
authoryear, 12pt]{elsarticle}
\usepackage[a4paper]{geometry}
\usepackage{amssymb}

\usepackage{changepage}
\usepackage{setspace}

\usepackage{chngcntr}
\usepackage{caption}
\usepackage{psfrag}
\usepackage{graphicx}
\usepackage{latexsym}
\usepackage{keyval}
\usepackage{ifthen}
\usepackage{moreverb}
\usepackage{gnuplottex}
\usepackage{upgreek}
\usepackage{subfig}
 \usepackage{amsmath}
\usepackage{appendix}
\usepackage{color}
\usepackage[usenames,dvipsnames]{xcolor}
\usepackage{enumerate}
\usepackage{wrapfig}
\usepackage{titlesec}
\titlelabel{\thetitle \quad}

\usepackage{ulem}

\usepackage[section]{placeins}	
\makeatletter
\AtBeginDocument{%
	\expandafter\renewcommand\expandafter\subsection\expandafter{%
		\expandafter\@fb@secFB\subsection
	}%
}
\makeatother

\makeatletter
\AtBeginDocument{%
	\expandafter\renewcommand\expandafter\subsubsection\expandafter{%
		\expandafter\@fb@secFB\subsubsection
	}%
}
\makeatother

\usepackage[hidelinks]{hyperref}

\newcommand{\eps}{\varepsilon}
\renewcommand{\d}{{\,\rm  d}}

\newcommand{\T}[1]{{#1}^{\sf  T}}

\newcommand{\pd}[2]{\displaystyle\frac{\partial #1}{\partial #2}}

\newcommand{\sign}[1]{{\rm sgn}\left( #1 \right)}
\newcommand{\grad}[1]{{\rm grad}\left( #1 \right)}
\newcommand{\bgrad}[1]{\overline{\rm grad}\left( #1 \right)}
\renewcommand{\div}[1]{{\rm div }\left( #1 \right)}
\newcommand{\bdiv}[1]{\overline{\rm div }\left( #1 \right)}

\newcommand{\fsym}[1]{{\rm sym }( #1 )}

\newcommand{\soutaa}[1]{}
\newcommand{\fempty}[1]{{}}
\newcommand{\f}[1]{\mbox{$ #1 $}}

\newcommand{\sty}[1]{\mbox{\boldmath $#1$}}

\newcommand{\styy}[1]{{\mathbb{#1}}}
\newcommand{\fa}{\sty{ a}}

\newcommand{\fd}{\sty{ d}}
\newcommand{\fe}{\sty{ e}}

\newcommand{\fl}{\sty{ l}}

\newcommand{\fn}{\sty{ n}}

\newcommand{\fr}{\sty{ r}}
\newcommand{\fs}{\sty{ s}}
\newcommand{\ft}{\sty{ t}}
\newcommand{\fu}{\sty{ u}}

\newcommand{\fx}{\sty{ x}}

\newcommand{\fzero}{\sty{ 0}}
\newcommand{\fA}{\sty{ A}}
\newcommand{\fB}{\sty{ B}}
\newcommand{\fC}{\sty{ C}}

\newcommand{\fE}{\sty{ E}}

\newcommand{\fG}{\sty{ G}}
\newcommand{\fH}{\sty{ H}}
\newcommand{\fI}{\sty{ I}}

\newcommand{\fK}{\sty{ K}}
\newcommand{\fL}{\sty{ L}}
\newcommand{\fM}{\sty{ M}}

\newcommand{\fP}{\sty{ P}}
\newcommand{\fQ}{\sty{ Q}}
\newcommand{\fR}{\sty{ R}}
\newcommand{\fS}{\sty{ S}}

\newcommand{\fU}{\sty{ U}}
\newcommand{\fV}{\sty{ V}}

\newcommand{\fX}{\sty{ X}}
\newcommand{\fY}{\sty{ Y}}
\newcommand{\fZ}{\sty{ Z}}

\newcommand{\ffC}{\styy{ C}}

\newcommand{\ffI}{\styy{ I}}

\newcommand{\ffR}{\styy{ R}}

\newcommand{\fsigma}{\mbox{\boldmath $\sigma$}}

\newcommand{\fomega}{\mbox{\boldmath $\omega$}}
\newcommand{\fOmega}{\mbox{\boldmath $\Omega$}}

\newcommand{\fxi}{\mbox{\boldmath $\xi $}}

\newcommand{\feps}{\mbox{\boldmath $\varepsilon $}}
\newcommand{\fEps}{\mbox{\boldmath $\mathcal{E} $}}
\newcommand{\fepsilon}{\mbox{\boldmath $\epsilon $}}

\newcommand{\fSigma}{\mbox{\boldmath $\Sigma $}}

\newcommand{\frho}{\mbox{\boldmath $\rho $}}

\newcommand{\cB}{{\cal B}}
\newcommand{\cC}{{\cal C}}

\newcommand{\cE}{{\cal E}}

\newcommand{\cZ}{{\cal Z}}
\definecolor{boxrahmen}{gray}{0.0}
\definecolor{boxhintergrund}{gray}{0.999}
\newsavebox{\tmpbox}

\newcommand{\mini}[1]{\underset{{#1}}{\rm min}}

\newboolean{badQuality}
\setboolean{badQuality}{false}

\RequirePackage{ifthen}
\newcommand{\blds}[1]{\mbox{\scriptsize \boldmath $#1$}}
\newcommand{\avg}[1]{\langle#1\rangle}
\newcommand{\und}[1]{\underline{#1}}

\newcommand{\phitlde}{\tilde \fu}

\newcommand{\bH}{\bar \feps}
\newcommand{\bB}{\bar \fsigma}
\newcommand{\tcH}{\tilde {\sty \cE}}
\newcommand{\tcEps}{\tilde {\sty \fEps}}
\newcommand{\bcEps}{\bar {\sty \fEps}}
\newcommand{\NipFE}{N_{\rm ip}^{\rm FE}}
\newcommand{\NipHR}{N_{\rm g}}
\newcommand{\Nmd}{N_{\rm md}}

\newcommand{\Nsim}{N_{\rm sim}}
\newcommand{\OmegaFE}{\Omega_{\rm FE}}

\newcommand{\bx}{\bar \fx}

\newcommand{\gammaacc}{\gamma^{\rm acc}}

\newcommand{\Nslp}{N_{\rm slp}}
\newcommand{\fepse}{\feps^{\rm e}}
\newcommand{\fepsp}{\feps^{\rm p}}
\newcommand{\Ms}{\fM^{\rm s}_\alpha}
\newcommand{\tauc}{\tau^{\rm c}}
\newcommand{\ttauc}{\tilde \tau^{\rm c}}
\newcommand{\tauD}{\tau^{\rm D}}
\newcommand{\csig}{c^{\bar \sigma}}
\newcommand{\cres}{c^R}
\newcommand{\peq}{+\hspace{-1mm}=}
\newcommand{\tfsigma}{\tilde \fsigma}

\begin{document}

\newlength{\smallDist} \newlength{\smallDistt}
\newlength{\smallDisttt} \newlength{\smallDistttt} 
\newlength{\vDist}


\begin{frontmatter}



\title{Computational Crystal Plasticity Homogenization using Empirically Corrected Cluster Cubature (E3C) Hyper-Reduction\\ (Preprint)}


\author[1]{Stephan Wulfinghoff}

\address[1]{Computational Materials Science, Department of Materials Science, Kiel University, Kaiserstr.~2, 24143 Kiel, Germany\\
swu@tf.uni-kiel.de}
\begin{abstract}
The computational homogenization of elastoplastic polycrystals is a challenging task due to the huge number of grains required, their complicated interactions and due to the complexity of crystal plasticity models per se. Despite a few successes of reduced order models, mean field and simplified homogenization approaches often remain the preferred choice. In this work, a recently proposed hyper-reduction method (called E3C) for projection-based Reduced Order Models (pROMs) is applied to the problem of computational homogenization of geometrically linearly deforming elastoplastic polycrystals. The main novelty lies in the identification of reduced modes (the 'E3C-modes'), which replace the strain modes of the reduced-order model, leading to a significantly smaller number of integration points. The peculiarity, which distinguishes the method from more conventional hyper-reduction techniques, is that the E3C integration points are not taken from the set of FE integration points. Instead, they can be interpreted as generalized integration points in strain space which are trained such as to satisfy an orthogonality condition, which ensures that the hyper-reduced model matches the equilibrium states and macroscopic stresses of full-field model data as accurately as possible. In addition, the number of grains is reduced, preserving the main features of the original texture of the finite element model. Two macroscopic engineering parts (untextured and textured) are simulated, illustrating the performance of the method in three-dimensional two-scale applications involving hundreds of thousands macroscopic degrees of freedom and millions of grains with computing times in the order of hours (cumulated online and offline effort) on standard laptop hardware.
\end{abstract}


\end{frontmatter}
\section{Introduction}
Metals are the material class of choice for countless applications in engineering fields like construction, transportation and beyond. During the design process of advanced mass products and costly individually produced parts, simulation tools for the optimization of the elastoplastic behavior have become indispensable in many cases. 
In addition, costly mass production tools and the production process itself are often subject to simulation-based optimization. For this purpose, phenomenological elastoplasticy models with varying complexity are readily available \citep[e.g.,][]{von1913mechanik, hill1948theory}. State-of-the-art models achieve excellent predictions of the initial yield surface \citep[][]{barlat2003plane, bron2004yield}, which comes at the cost of an increased number of material parameters and the need for extended experiments for the parameter calibration. The prediction of the yield surface evolution is significantly more challenging, in particular when the loading is non-proportional \citep{pietryga2012finite}. Furthermore, any material modifications concerning the microstructure, like grain size, second phases, or texture, require an expensive recalibration of the model parameters. For this reason, it is difficult to develop a concurrent design strategy, aiming at a simultaneous optimization of macroscopic and microscopic properties, using exclusively phenomenological macroscopic models. The same holds for macroscopic property heterogeneities, like surface hardened zones, which are often intentionally introduced during the production process. Two-scale models resolve both, the component scale as well as the microstructure \citep{benedetti2013modelling}, and come with the promise to not only require less parameter recalibration upon microstructural modifications but, on the contrary, even predict their impact on the component's mechanical behavior \citep{segurado2018computational}. For metals, the material models of choice originate from the field of continuum crystal plasticity, with ground-breaking contributions by \cite{hill1966generalized} (small deformations), \citet{rice1971inelastic} as well as \citet{hill1972constitutive} (large deformations). This constitutive framework is nowadays widely applied with well established models for viscoplasticity \citep{peirce1983material}, for isotropic hardening \citep{peirce1982analysis, franciosi1982multislip, kocks2003physics} and for kinematic hardening \citep{me1991single}. The numerical time integration of crystal plasticity models equations is challenging \citep{needleman1985finite, cuitino1993computational} and, with limited computational resources, it took many years until realistic three-dimensional finite element simulations of polycrystalline microstructures became feasible \citep{barbe2001intergranular, barbe2001intergranularb}. The Fast Fourier Transform method \citep{moulinec1994fast} is a very efficient alternative, and has been successfully applied to numerous polycrystalline microstructures  \citep{lebensohn2001n}. These full-field simulations showed that {\it i}) representative volume elements (RVEs) with hundreds up to thousands of crystals are usually required in order to provide reliable homogenized properties and that {\it ii}) the local stress and strain fields can fluctuate significantly within a single grain and the grain-grain interactions play an important role in the microscopic deformation processes \citep{barbe2001intergranular, barbe2001intergranularb}. The tremendous computational effort of full-field models limits their direct application in numerical two-scale frameworks, like FE\f{^2}- or FE-FFT-methods, to comparably simple or two-dimensional problems \citep[e.g.,][]{miehe1999computational, kochmann2017efficient}. To date, more traditional mean field homogenization methods usually remain the preferred two-scale simulation strategy, not least because they are particularly well suited for the specific material class of polycrystals. Even the comparably simple Taylor assumption \citep{taylor1938plastic} leads to reasonable predictions of the macroscopic stress strain response and texture evolution, but requires a large number of crystal orientations, as the macroscopic anisotropy is otherwise overpredicted. For this reason, \citet{bohlke2005texture} approximated the crystal orientation distribution function via texture components modeled by Mises-Fischer distributions, allowing to significantly reduce the number of required single crystal orientations and to carry out complex macro-scale simulations \citep{bohlke2006finite}.\\
A very popular mean field method for polycrystals is the self-consistent scheme, which was originally applied to elastic polycrystals \citep{hershey1954elasticity, kroner1958berechnung} and is under certain conditions equivalent to the Hashin-Shtrikman variational procedure \citep{hashin1962some,hashin1962variational,hashin1963variational}. The generalization to viscoplasticity \citep{molinari1987self, lebensohn1993self} leads to the so-called viscoplastic self-consistent approximation and requires a grain-wise (tangent- or secant-based) linearization of the nonlinear material law, the optimal choice of which is still subject of ongoing research \citep{lebensohn2007self, song2018fully}. A simplified self-consistent scheme is used in the Reduced Texture Methodology (RTM) introduced by \citet{rousselier2006simplified}, which relies on a very small number of (in the order of ten) representative grains with a low computational cost. Instead of deriving the crystal orientations of the model from crystallographic measurements, they are treated like phenomenological model parameters, i.e., they are fitted such that the macroscopic stress response matches experimentally obtained stress strain data, with surprisingly good results. The model successfully captures important macroscopic features of anisotropic plasticity under non-proportional loading and continues to be applied and further developed \citep{rousselier2009novel,rousselier2012macroscopic,kong2023polycrystalline}.\\
Besides its widespread use in the context of mean field modeling, the self-consistent scheme also has an impact on model order reduction methods with a higher resolution of the micro-fields and intragranular fluctuations. One example is the self-consistent clustering analysis, proposed by \citet{liu2016self} and applied to polycrystals undergoing finite deformations by \citet{yu2019self}, which is largely equivalent to the Hashin-Shtrikman type finite element method \citep{wulfinghoff2018model, cavaliere2020efficient}. Most other model order reduction (MOR) methods show little commonalities with the approximations made in mean field theories, but have proven to be powerful alternatives \citep{yvonnet2007reduced, hernandez2014high}. Most MOR approaches define a reduced order model (ROM) in terms of global shape functions -- so-called 'modes' -- being derived from on a collection of full-field simulation data (the 'snapshots') via proper orthogonal decomposition \citep[POD, e.g.,][]{holmes1996turbulence}. It turns out that the solution indeed resides in a low-dimensional subspace, such that 10-100 of these modes are often sufficient to capture the full richness of the microscopic strain and stress fields. The related degrees of freedom are the mode coefficients, which are obtained via the classical Galerkin (or Petrov-Galerkin) method, i.e., the high-dimensional residual is projected onto the low-dimensional reduced solution space (hence the name 'projection-based ROMs' or pROMs). The involved nonlinear residuals are evaluated by a costly numerical integration procedure, requiring a huge number of nonlinear computations over the whole microstructure.\\
Tremendous further speed-ups can be achieved via 'hyper-reduction' methods \citep{ryckelynck2009hyper}, which aim at a significantly accelerated evaluation of the nonlinear integrals, making use of the fact that the aforementioned integrands can also be approximated in a low dimensional space \citep[see the recent overview by ][]{bhattacharyya2025hyper}. Most of such hyper-reduced order models (HROMs) can be classified into two categories -- the 'approximate-then-project' and the 'project-then-approximate' types. During the offline phase, approximate-then-project methods derive modes also for the nonlinear integrands from the high-fidelity snapshots. During the online phase, they evaluate the nonlinear integrand on a small subgrid (e.g., a small subset of the entirety of finite elements) and identify that linear combination of modes which best fits the subgrid values. Such 'gappy-POD' approaches (and the like) find their origins in image reconstruction \citep{everson1995karhunen}. The obtained full-field approximation of the integrand can finally be Galerkin-projected onto the solution subspace (these steps are in fact condensed into one small matrix online operator), leading to a reduced nonlinear equation system for the mode coefficients. Important approximate-then-project methods include, but are not limited to, the Discrete Empirical Interpolation Method \citep[DEIM, or discrete EIM, see][]{barrault2004empirical,chaturantabut2009discrete}, different variants thereof \citep[e.g.,][]{peherstorfer2014localized, lauzon2024s} and the Gauss-Newton method with Approximated Tensors \citep[GNAT,][]{carlberg2011efficient,carlberg2013gnat}, which minizes the squared residual in a Petrov-Galerkin sense.\\
While approximate-then-project methods may lack robustness under certain circumstances \citep[see the surveys by][]{van2018integration, brands2019reduced}, project-then-approximate inherit the convexity-properties -- if present -- of the underlying full-field model, and they are therefore often preferred \citep{lange2024monolithic}. The Energy Conserving Sampling and Weighting method (ECSW), developed by \citet{farhat2014dimensional}, essentially underintegrates the nonlinear terms of a pROM obtained through Galerkin projection of a nonlinear finite element model. To this end, the assembly process is performed on a subgrid, choosing the weights of the few remaining element contributions such as to minimize the approximation error in the Galerkin-projected residual. The Empirical Cubature Method (ECM) by \citet{hernandez2017dimensional} further refines this approach in selecting individual integration points rather than whole elements. The authors make use of a truncated singular value decomposition of the integrand snapshot matrix to efficiently compute the reduced integration weights. The continuous ECM (CECM) generalizes this concept by extending the search for the reduced integration points from the Gauss points to the whole computational domain \citep{hernandez2024cecm}, effectively replacing the previous, hard-to-solve combinatorial task (selection of points) by a smoother problem, enabling the application of the Newton method.\\
One of the rare applications of model order reduction methods to the computational homogenization of elastoplastic polycrystals was effectuated by \citet{zhang2015eigenstrain}. This eigenstrain based approach was influenced by the Transformation Field Analysis (TFA) of \citet{dvorak1992transformation} and was generalized to large deformations by \citet{xia2021large}. Another model order reduction approach to polycrystals was developed by \citet{michel2016model,michel2016poly}, who combined their Nonuniform Transformation Field Analysis \citep[NTFA,][]{michel2003nonuniform} with a Tangent Second Order approximation \citep[TSO,][]{castaneda1996exact}, using a mixed form due to \citet{fritzen2013reduced}. This combination leads to a homogenized generalized standard material \citep{halphen1975generalized} with plastic modes on the microscale and grain-wise linearized constitutive equations, accurately accounting for the impact of the plastic fluctuations on the macroscopic stress response. The NTFA-TSO framework combines treatmendous speed-ups with a high accuracy and has recently been applied by \citet{labat2023multiscale} to model the behaviour of polycrystalline uranium dioxide fuel.\\
One argument to step back from the tangent-based linearization of the constitutive equations was given by \citet{ponte2015fully, castaneda2016stationary} (and references therein), who promotes a generalized secant approximation as a promising homogenization approach. Simply speaking, in case of stronly nonlinear microstructures and intense field fluctuations, more accurate results can be expected, if the effective stress within a certain phase (e.g., a grain) is averaged over the stress response of a small set of phase-specific strain values, rather than to use a linearization of the material law around the phase's strain average \citep[see Eq.~(49) in][for the dual statement]{castaneda2016stationary}. This set of strain values should be representative of the strain field within the respective phase. In the aforementioned reference, these strain values were taken to be statistically representative of the actual microscopic strain field in the sense that the first two statistical moments are exactly matched. This idea was transferred (and generalized) to the field of reduced order modeling by \citet{michel2017effective} for the special case of crystal plasticity, using the NTFA-framework, and by \citet{wulfinghoff2024statistically} to speed up Galerkin projection based ROMs. Both types of reduced order models use global shape functions / modes with the advantage of a cheap online computation of the aforementioned statistically representative strain values. A key feature of these approaches is that these strain values, used for the material law evaluation, replace the integration points of the full-field model, i.e., the {\it they are not taken as a subset thereof} (in contrast to most hyper-reduction approaches). For this reason, they were called 'generalized integration points' by \citet{wulfinghoff2024statistically}. These integration points exactly preserve the field statistics up to second order \citep[or even higher, see][]{michel2017effective}. Furthermore, they preserve the mode character of the reduced-order model, i.e., for every mode of the fully integrated reduced order model, there is also a mode in terms of the generalized integration points. This feature enables a fast reconstruction of the full-field solution, in a post processing step.
\begin{figure}[h]
\centering
  \includegraphics[width=0.99\textwidth]{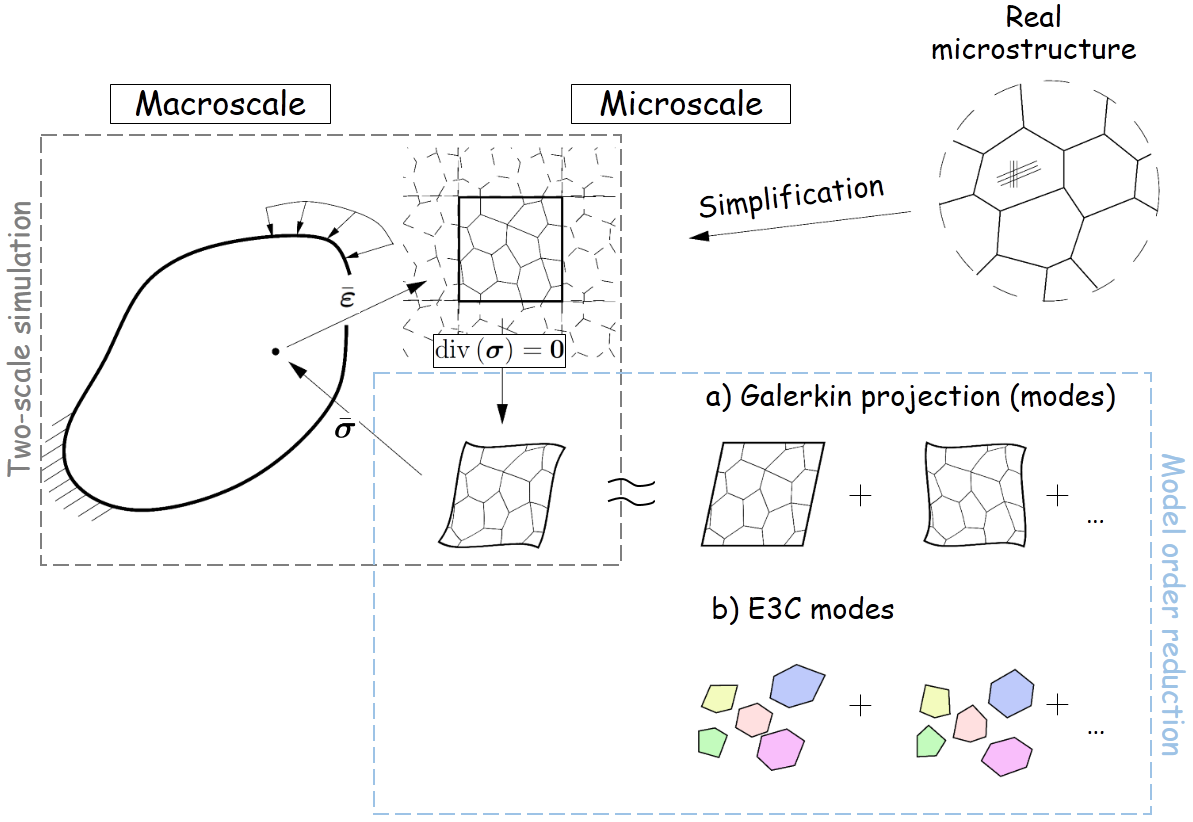}
\caption{Sketch of the overall computational homogenization scheme.}
\label{handgemalt}
\end{figure}\\
The strategy followed in this work is based on the hypothesis that 'reduced modes' in terms of statistically representative (generalized) integration points yield good predictions, but are not necessarily the optimal choice, if the macroscopic response is the main quantity of interest. Instead, the idea is to identify reduced modes, which are optimal in the sense that they correctly predict both, the mode coefficients (required for the full-field reconstruction) and the macroscopic stresses. Instead of relying on the field statististic, this is achieved by machine learning, i.e., by plain minimization of a related cost function, which measures the deviation of the reduced-order model from full-field data. The resulting hyper-reduction method is called Empirically Corrected Cluster Cubature (E3C) and was introduced by \citet{wulfinghoff2025empirically} for two-dimensional magnetostatic computational homogenization. It was transferred to infinitesimal and finite nonlinear elasticity by \citet{wulfinghoff2025e3c} and \citet{wulfinghoff2025homogenization}, respectively. The first three-dimensional E3C-application was provided in the context of magnetostatic computational homogenization by \citet{goldbeck2025computational}.\\
The overall procedure is illustrated in Fig.~\ref{handgemalt}\footnote{Figure and text adapted from \citet{wulfinghoff2025homogenization}.}. The upper left part of the figure depicts the usual first order homogenization approach, i.e., the macroscopic strain~\f{\bar \feps} is passed to the microscale, where a crystal plasticity model is applied to a set of grains with varying crystallographic orientations. The microscopic deformation is then computed by solving the micro equilibrium equations in combination with some suitable (e.g., periodic) boundary conditions and subsequently, the average stress is passed back to the macroscale.
\\
The model order reduction approach (bottom, right) comprehends two parts: a) The Galerkin projection, where the microscopic deformation fields are represented in terms of modes (i.e., global shape functions, see Fig.~\ref{handgemalt}) and b) E3C hyper-reduction. In this work, the hyper-reduction step involves the identification of reduced modes, i.e., the crystal plasticity material law is evaluated only once per grain. Furthermore, the full granular microstructure is replaced by a reduced set of grains, in order to minimize the computational cost as much as possible.\vspace{3mm}\\
{\bf Notation.} The symmetric part of a 2nd-order tensor is denoted as \f{\fsym{\fC}=(\fC+\T{\fC})/2}. 
The macroscopic gradient of a vector field~\f{\bar \fa(\bx)} reads \f{\bgrad{\bar \fa}=\partial \bar a_i/\partial \bar x_j \fe_i \otimes \fe_j} with orthonormal base vectors~\f{\fe_i}. The double contraction of two second-order tensors is designated by \f{\fA:\fB=A_{ij}B_{ij}}.  
The third-order permutation tensor is denoted as \f{\fepsilon}.
\begin{figure}
\centering
\framebox{
\begin{minipage}{.9\textwidth}
\vspace{3mm}
 \begin{itemize}
  \item Macro-problem (neglecting inertia and body forces)
  \begin{itemize}
   \item Strain (kinematics)
   \begin{equation}
    \bar \feps=\fsym{\bgrad{\bar \fu}} \ \rm{in}\  \bar \cB
   \end{equation}
   \item Linear momentum balance
   \begin{equation}
    \bdiv{\bar \fsigma} = \fzero \ \rm{in}\  \bar \cB
   \end{equation}
   \item Boundary conditions
   \begin{equation}
    \bar \fsigma \bar \fn = \bar\ft_0 \ \rm{on}\ \partial \bar \cB_t,\ \bar \fu = \bar \fu_0\ \rm{on}\ \partial \bar \cB_{\rm{u}}
   \end{equation}
  \end{itemize}
%
  \item Micro-problem
  \begin{itemize}
   \item Kinematics
   \begin{align}
    &\fu(\bx, \fx,t)=\bar \feps(\bar \fx,t) \fx + \tilde \fu(\fx,t), \  \rm{in}\ \Omega, \ \tilde \fu \ {\rm periodic\ on}\ \partial \Omega \\
    &\feps=\fsym{\grad{\fu}}=\bar \feps+\tilde \feps \ \rm{in}\ \Omega 
   \end{align}
   \item Linear momentum balace
   \begin{equation}
    \div{\fsigma} = \fzero \ \rm{in}\  \Omega \label{microlmb}
   \end{equation}
    + constitutive law (see text)
  \end{itemize}
%
  \item Scale transition
  \begin{flalign}
   &\bar \feps = \langle \feps \rangle, \ \bar \fsigma = \langle \fsigma \rangle \ {\rm with} \ \avg{\bullet}=\frac{1}{\Omega}\int_\Omega \bullet \d \Omega&& \label{barqtts}
  \end{flalign}

 \end{itemize}
 \vspace{1mm}
\end{minipage}
}
\caption*{Box~1: Summary of two-scale problem: \f{\fu}: displacement, \f{\feps}: strain tensor, \f{\fsigma}: stress tensor, \f{\ft}: traction vector, \f{(\bar \bullet)} macroscopic quantity, \f{(\tilde \bullet)} fluctuation, \f{(\bullet)_0}: given quantity, \f{\bar \cB}: solid (macroscale), \f{\Omega}: microstructure.}
\end{figure}
\section{Two-scale problem}\label{muBVP}
This work is concerned with the geometrically linear deformation of statistically homogeneous polycrystalline solids, to be approximated by conventional first-order homogenization theory. For the theoretical fundations see, e.g., \citet{hill1963elastic}, \citet{bensoussan1978asymptotic} or \citet{sanchez1980non} and for early computational realizations (FE\f{^2}) approaches, refer to \citet{smit1998prediction} or \citet{miehe1999computational}, amongst many others. Within this framework, a macroscopic initial boundary value problem is to be solved, where the material stress strain relation is obtained through the averaged response of a micro-problem, being attached to each macroscopic material point. The corresponding theory is well document in the cited literature and is summarized in Box~1, for convenience.
The weak form of the microscopic linear momentum balance (Eq.~\eqref{microlmb}) reads\footnote{Recall the relation to the Hill-Mandel lemma: \f{\avg{\fsigma:\delta \feps}=\avg{\fsigma:\delta \bar \feps}+\avg{\fsigma:\delta \tilde\feps} \stackrel{\eqref{barqtts}, \eqref{weakform}}{=}\bar \fsigma:\delta\bar\feps}.}
\begin{equation}
 \avg{\fsigma:\delta \tilde \feps}= \frac{1}{\Omega} \int \limits_\Omega \fsigma : \delta \tilde \feps \d \Omega = 0\label{weakform}
\end{equation}
with the variation of the microscopic strain fluctuation being given by \f{\delta \tilde \feps=\fsym{\grad{\delta \tilde \fu}}}.
\section{Crystal plasticity model}
\subsection{Constitutive equations}
The microstructure \f{\Omega} is assumed to be given by a periodic polycrystal, being composed of a finite number of idealy interconnected grains. The latter are characterized by individual lattice orientations, as described by the crystallographic slip plane normals \f{\fn_\alpha=\fQ\fn_{\alpha 0}} and slip directions \f{\fd_\alpha=\fQ \fd_{\alpha 0}}, where \f{\alpha \in (1,\dots,\Nslp)} is the slip system index, \f{\fn_{\alpha 0}} and \f{\fd_{\alpha 0}} are orthonormal reference vectors and \f{\fQ} is a proper orthogonal tensor, describing the orientation of a specific crystal with respect to the reference frame. The visco-elastoplastic constitutive model equations are summarized in Box~2. They comply with the standard literature \citep[e.g.,][]{hill1966generalized, teodosiu1970dynamic, rice1971inelastic} and use an isotropic hardening model based on the accumulated plastic slip \f{\gammaacc}, treating self- and cross-hardening equally, for simplicity. In this work, the rate sensitivity exponent is chosen as \f{p=20}, in order to approximate rate-independent behavior with a critical resolved shear stress being given by \f{\tauD+\ttauc(\gammaacc)}. The internal variables \f{\fX=\{\fepsp,\gammaacc\}} are initialized as zero.
\section{Model order reduction}
\subsection{Galerkin projection}
\begin{figure}
\centering
\framebox{
\begin{minipage}{.9\textwidth}
\vspace{3mm}
 \begin{itemize}
  \item Kinematics
  \begin{flalign}
    &\feps=\fepse+\fepsp, \ \ \ \fepsp=\sum\limits_{\alpha=1}^{\Nslp} \gamma_\alpha \Ms, \ \ \ \Ms=\fsym{\fd_\alpha \otimes \fn_\alpha} \label{kinematics}
  \end{flalign}
  \item Hooke's law and resolved shear stresses
  \begin{flalign}
    &\fsigma=\ffC:\fepse, \ \ \ \tau_\alpha = \fsigma : \Ms \label{Hookeslawandshearstresses}
  \end{flalign}  
  \item Flow rule
  \begin{flalign}
   &\dot \gamma_\alpha = \sign{\tau_\alpha} \dot \gamma_0 \left\langle \frac{|\tau_\alpha|-\tauc}{\tauD} \right\rangle^p \label{flowrule}
  \end{flalign}
  \item Hardening law
  \begin{flalign}
   &\tauc = \ttauc(\gammaacc)= q \left( \frac{\gammaacc}{\gamma_0} \right)^n, \ \ \ \dot \gamma^{\rm acc}=\sum\limits_{\alpha=1}^{\Nslp} |\dot \gamma_\alpha| \label{hardeninglaw}
  \end{flalign}
 \end{itemize}
 \vspace{1mm}
\end{minipage}
}
\caption*{Box~2: Crystal plasticity model: \f{\fepse}: elastic strain, \f{\fepsp}: plastic strain, \f{\gamma_\alpha}: plastic slips, \f{\Ms}: symmetric part of Schmid tensor, \f{\ffC}: stiffness/elasticity tensor, \f{\dot \gamma_0}: reference shear rate, \f{\tauc}: critical resolved shear stress, \f{\tauD}: drag stress, \f{p}: rate sensitivity exponent, \f{q, \gamma_0, n}: hardening parameters.}
\end{figure}
The numerical solution of the microscopic problem, outlined in the previous sections, relies on a discretization of the displacement fluctuation in terms of global shape functions \f{\tilde \fU_k(\fx)}:
\begin{equation}
 \phitlde(\fx,t) = \sum\limits_{k=1}^{\Nmd} \xi_k(t) \tilde \fU_k(\fx).
\end{equation}
The shape functions \f{\tilde \fU_k(\fx)} are called modes, and they are assumed given\footnote{The \f{\tilde \fU_k(\fx)} are assumed to be normalized. They may be obtained from a representative set of finite-element simulations (snapshots) via proper orthogonal decomposition. This standard approach is well documented in the cited literature and is not repeated here.}. The primary unknowns are the mode coefficients \f{\fxi=(\xi_1,\dots,\xi_{\Nmd})}. This discretization implies the existence of strain fluctuation modes 
\begin{equation}
\tcH_k(\fx)={\rm sym}\left(\grad{\tilde \fU_k(\fx)}\right), \label{eqstrainmodes}
\end{equation}
 and the microscopic strain takes the form
\begin{equation}
 \feps(\fx,t) = \bar \feps(t) + \tilde \feps(\fx,t) = \bH(t) + \underbrace{\sum\limits_{k=1}^{\Nmd} \xi_k(t) \tcH_k(\fx)}_{\tilde {\blds \varepsilon}({\blds x},t)}.\label{microstrain}
\end{equation}
Here and in the following, the dependence on the macroscopic position \f{\bx} is dropped, for notational simplicity.
It is assumed that the strain variation is discretized in analogy to the strain, such that insertion into the weak form (see Eq.~\eqref{weakform}) yields:
\begin{align}
 \frac{1}{\Omega}\int \limits_\Omega \fsigma : \delta \tilde \feps \d \Omega=&\sum\limits_{k=1}^{\Nmd} \delta \xi_k \, \frac{1}{\Omega}\int \limits_\Omega \tcH_k(\fx)  :  \fsigma(\fx,\feps,\fX) \d \Omega \nonumber \\
 \approx &\sum\limits_{k=1}^{\Nmd} \delta \xi_k \underbrace{ \frac{1}{\Omega}\left( \sum\limits_{p=1}^{\NipFE} \tcH_k(\fx^p)  :  \fsigma\big(\fx^p,\feps(\fx^p),\fX(\fx^p)\big) \, \OmegaFE^p\right)}_{=:R_k} \nonumber\\
 =&\delta \fxi  \cdot  \fR=0. \label{discweakform}
\end{align}
Here, the \f{\fx}-dependence in \f{\fsigma(\fx,\feps,\fX)} accounts for the heterogeneity of the microstructure, i.e., each grain exhibits its individual crystallographic orientation. The residuals \f{R_k} are evaluated based on a further approximation of the integrals in Eq.~\eqref{discweakform} by an appropriate quadrature rule. In this work, it is assumed that a finite element (FE) model of the microstructure is available. For the reduced-order model, one integration point \f{\fx^p} per element \f{\OmegaFE^p} (\f{p=1,\dots,\NipFE}) is used, using element-wise averaged strains / B-matrices\footnote{In contrast to the finite element model, reduced integration is not expected to be detrimental, since hourglass instabilities are not encoded in the modes of the reduced-order model.}.\\
The residual vector \f{\und{R}} in Eq.~\eqref{discweakform} must vanish due to the arbitraryness of \f{\delta \fxi}:
\begin{equation}
 \fR = \begin{pmatrix} R_1\\ \vdots \\ R_{\Nmd} \end{pmatrix} = \fzero.\label{resid}
\end{equation}
Equation~\eqref{resid} describes a set of \f{\Nmd} nonlinear equations for the unknown mode coefficients~\f{\fxi}, which must be solved in combination with the constitutive equations (Box~2) at the integration points. The macroscopic stress \f{\bB} is defined by
\begin{equation}
 \bB= \avg{\fsigma}= \frac{1}{\Omega} \int\limits_\Omega \fsigma \d \Omega \approx \frac{1}{\Omega} \sum\limits_{p=1}^{\NipFE} \fsigma(\fx^p,\feps(\fx^p),\fX(\fx^p)) \, \OmegaFE^p. \label{Bbar}
\end{equation}
\subsection{Empirically Corrected Cluster Cubature (E3C)}
\subsubsection{General hyper-reduction}
The bottleneck of the Galerkin projection method, as described in the previous section, is the assembly of the residual vector~\f{\fR} and the stiffness matrix \f{\fK=\partial \fR/\partial \fxi} (after time discretization), which requires a large number of expensive crystal plasticity computations (one per finite element). The aim of hyper-reduction methods, like the E3C-method presented below, is to substantially reduce this computational cost, based on a strongly reduced number \f{\NipHR \ll \NipFE} of material law evaluations. Correspondingly, the hyper-reduced versions of Eqns.~\eqref{discweakform} and \eqref{Bbar} read
\begin{equation}
 R_k \approx \frac{1}{\Omega}\sum\limits_{q=1}^{\NipHR} \tcH_k^q  :  \fsigma^q \, \Omega^q=0,  \ \ \ \bB \approx \frac{1}{\Omega} \sum\limits_{q=1}^{\NipHR} \fsigma^q \, \Omega^q\label{apprs}.
\end{equation}
Equation \eqref{apprs} now requires a significantly reduced number of stress computations. The weights~\f{\Omega^q} differ from those of the FE model. The quantities \f{\Omega^q} and \f{\tcH_k^q} are yet to be defined. In a time-discrete setting, the stress \f{\fsigma^q} and strain \f{\feps^q} in Eq.~\eqref{apprs} may be expressed formally as 
\begin{equation}
  \fsigma^q_{n+1}=\fsigma(\feps^q_{n+1}, \fX^q_n, \fQ^q), \ \ \ \feps^q_{n+1}=\bar \feps_{n+1} + \tilde \feps^q_{n+1}                \label{sigmaq}
\end{equation}
in the sense that the stress at the end of a given time step \f{n \to n+1} can be computed from the strain \f{\feps^q_{n+1}} at the end of the step and the internal variables \f{\fX^q_n} at the beginning of the step. In addition, the rotation tensor \f{\fQ^q}, describing the crystal orientation, is explicitly listed as argument. The reason for this choice will become clear below.
\subsubsection{E3C-kinematics}
\begin{figure}[h]
\centering
  \psfrag{q1}{\scriptsize \textcolor{BlueViolet}{\f{q=1}}}
  \psfrag{q2}{\scriptsize \textcolor{RoyalBlue}{\f{q=2}}}
  \psfrag{q3}{\scriptsize \textcolor{CadetBlue}{\f{q=3}}}
  \psfrag{q4}{\scriptsize \textcolor{Gray}{\f{q=4}}}
  \psfrag{q5}{\scriptsize \textcolor{Tan}{\f{q=5}}}
  \psfrag{q6}{\scriptsize \textcolor{Bittersweet}{\f{q=6}}}
  \psfrag{q7}{\scriptsize \textcolor{BrickRed}{\f{q=7}}}
  \psfrag{p0}{\hspace{-10mm}\f{\tilde \fE=\xi_1}}
  \psfrag{p1}{\f{+ \ \xi_2}}
  \psfrag{p2}{\f{+ \ \xi_3}}
  \psfrag{p3}{\f{+\ \xi_4}}
  \psfrag{...}{\f{+\dots}}
  \psfrag{un}{\footnotesize undeformed}
  \includegraphics[width=0.9\textwidth]{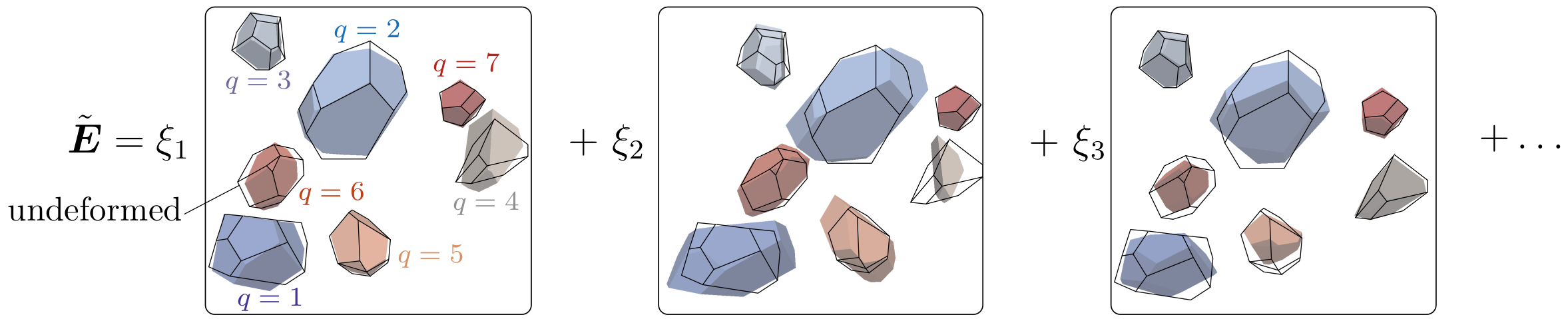}
\caption{Tentative visualization of the E3C-modes \f{\tcEps_k} (columns of matrix \f{\tcEps}) for the special case of seven E3C-grains (\f{\NipHR=7}). Black lines: undeformed grains, colored: deformed grains. The grains are disconnected due to the incompatibility of the grain-wise constant strains.}
\label{E3Cmodes}
\end{figure}

In the context of polycrystalline microstructures, the aim of the E3C-method is to replace the numerous grains of the finite element model by a reduced set of \f{\NipHR} grains with volumes~\f{\Omega^q} and grain-wise constant strains \f{\feps^q=\bar \feps + \tilde \feps^q} (\f{q=1,\dots,\NipHR}). In principle, one may also introduce several strain fluctuation values \f{\tilde \feps^q} per grain, but this idea is not followed here due to the observations made by \citet{kochmann2018efficient}. The fluctuations \f{\tilde \feps^q} are constrained to vanish in average:
\begin{equation}
 \langle \tilde \feps \rangle = \frac{1}{\Omega} \sum \limits_{q=1}^{\NipHR} \tilde \feps^q \Omega^q = \fzero. \label{tldeepsavgzero}
\end{equation}
In analogy to Eq.~\eqref{microstrain}, the strain fluctuations are expressed as linear functions of the mode coefficients:
\begin{equation}
\tilde \feps^q=\sum\limits_{k=1}^{\Nmd}\xi_k \tcEps^q_k
\ \ \ \Leftrightarrow \ \ \ 
\tilde \fE:=
 \begin{pmatrix}
  \tilde \feps^1 \\ \tilde \feps^2 \\ \vdots \\ \tilde \feps^{\NipHR} 
 \end{pmatrix}
 =
 \underbrace{
 \left(\fbox{ \f{{{\tcEps^1_{1_{\textcolor{white}{1}}}}\atop{\tcEps^2_{1_{\textcolor{white}{1}}}}}\atop{{\vdots}\atop{\tcEps^{\NipHR}_{1_{\textcolor{white}{1}}}}}}} \
 \fbox{ \f{{{\tcEps^1_{2_{\textcolor{white}{1}}}}\atop{\tcEps^2_{2_{\textcolor{white}{1}}}}}\atop{{\vdots}\atop{\tcEps^{\NipHR}_{2_{\textcolor{white}{1}}}}}}} \ 
 \dots \ 
 \fbox{ \f{{{\tcEps^1_{\Nmd}}\atop{\tcEps^2_{\Nmd}}}\atop{{\vdots}\atop{\tcEps^{\NipHR}_{\Nmd}}}}} \right)}_{\tcEps}
  \begin{pmatrix}
  \xi_1 \\ \xi_2 \\ \vdots \\ \xi_{\Nmd}
 \end{pmatrix}. \label{tildeeps}
\end{equation}
The columns \f{\tcEps_k} (\f{k=1,\dots,\Nmd}, framed in Eq.~\eqref{tildeeps}) of the matrix \f{\tcEps} are the E3C strain fluctuation modes  (using Mandel notation\footnote{Mandel notation: \f{\fsigma=\T{(\sigma_{11},\sigma_{22},\sigma_{33},\sqrt{2}\sigma_{23},\sqrt{2}\sigma_{13},\sqrt{2}\sigma_{12})}}}), that remain to be defined. The individual entries \f{\tcEps^q_k} (\f{q=1,\dots,\NipHR}) of the \f{k}th E3C-mode \f{\tcEps_k} are defined grain-wise and replace the 'continuum' modes \f{\tcEps_k(\fx)}, which are fields. Note that the same symbol '\f{\tcEps_k}' is used for the columns of the matrix \f{\tcEps} and for the full modes, to emphasize their common meaning. They are uniquely distinguishable by consistently marking the continuum modes \f{\tcEps_k(\fx)} by their \f{\fx}-dependence. A graphical interpretation the E3C-modes is difficult, since the E3C-grains do not have a shape (or the shape may be chosen arbitrarily). Nevertheless, a tentative visualization of Eq.~\eqref{tildeeps} is given in Fig.~\ref{E3Cmodes}, where the first three deformation modes are illustrated for \f{\NipHR=7} exemplary E3C-grains. Their sum (weighted by the mode coefficients) results in the grain-wise constant strain fluctuations \f{\tilde \feps^q}.
\\
An important property of the model is that the fluctuations \f{\tilde \feps^q} are kinematically not independent. Instead, the grains can only deform collectively, as described by Eq.~\eqref{tildeeps} in terms of the modes. In other words, the vector \f{\tilde \fE} in Eq.~\eqref{tildeeps}, which has \f{6\NipHR} components, is constrained to lie in a \f{\Nmd}-dimensional subspace of \f{\ffR^{6\NipHR}}. This feature mimics the presence of the microstructure in combination with the compatibility of the strain field.\\
For later use, the vector \f{\bar \fE = \T{(\T{\bar \feps}, \T{\bar \feps}, \dots, \T{\bar \feps})}} is also defined, such that
\begin{equation}
 \fE=\begin{pmatrix}
  \feps^1 \\ \feps^2 \\ \vdots \\ \feps^{\NipHR} 
 \end{pmatrix} = \bar \fE + \tilde \fE 
 = \underbrace{{\footnotesize
 \begin{pmatrix}
   1 & 0 & \dots & 0\\
   0 & 1 & \dots & 0\\
   0 & 0 & \dots & 0\\
   0 & 0 & \dots & 0\\
   0 & 0 & \dots & 0\\
   0 & 0 & \dots & 1\\
   \vdots & \vdots & \ddots & \vdots \\
   0 & 0 & \dots & 1
  \end{pmatrix}} }_{\bcEps} \bar \feps + \tcEps \fxi. \label{Edecomp}
\end{equation}
The constraint 
\begin{equation}
 \avg{\tcEps_k} = \frac{1}{\Omega}\sum\limits_{q=1}^{\NipHR} \tcEps^q_k \Omega^q = \fzero \label{epscontst}
\end{equation}
ensures satisfaction of Eq.~\eqref{tldeepsavgzero}.
\subsubsection{Interpretation as cubature method in strain space}
A distinct feature of the E3C-method is that the upper index '\f{q}' of the E3C-mode entries~\f{\tcEps_k^q} does not correspond to any integration point \f{\fx^p} of the underlying finite element model. Nevertheless, the method allows for an interpretation as efficient cubature method. However, the integration does not take place in real space but in the six-dimensional strain space, where the strain distribution is characterized by a normalized probability density function \f{\rho(\feps)} with
\begin{equation}
 \int\limits_{\ffR^6} \rho(\feps)\d E=1, 
\end{equation}
where \f{\d E} is a six-dimensional hyper-volume element\footnote{Strictly, speaking this implies that the strain tensor is represented as a vector in \f{\ffR^6}, using, e.g., Voigt- or Mandel-notation.}. This statistical description of the strain field proves particularly useful in the context of representative volume elements of random composites \citep{kanit2003determination}. Such RVEs need to be sufficiently large in order to include a statistically relevant number of microscopic entities (e.g., hundreds of grains, inclusions or pores), leading to overwhelmingly complex stress and strain fields, which are often more tractable by a statistical description.\\
A corresponding finite element model, with discrete integration point values \f{\feps^p=\feps(\fx^p)}, is characterized by the density
\begin{equation}
 \rho(\feps)= \frac{1}{\Omega} \sum\limits_{p=1}^{\NipFE} \delta (\feps^p-\feps) \OmegaFE^p, 
\end{equation}
where \f{\delta(\feps)} denotes the Dirac delta function in strain space. In this case, the macroscopic strain is given by
\begin{equation}
 \bar \feps=\int\limits_{\ffR^6} \rho(\feps) \feps \d E=\frac{1}{\Omega} \int\limits_{\ffR^6} \sum\limits_{p=1}^{\NipFE}  \delta(\feps-\feps^p) \feps^p
  \OmegaFE^p \d E=\frac{1}{\Omega}  \sum\limits_{p=1}^{\NipFE} \feps^p \OmegaFE^p. \label{rhodelta}
\end{equation}
The macroscopic stress correspondingly reads
\begin{equation}
 \bar \fsigma = \int\limits_{\ffR^6} \frho_\sigma(\feps) \d E = \frac{1}{\Omega}  \sum\limits_{p=1}^{\NipFE} \fsigma^p \OmegaFE^p \ \ \ 
  {\rm with} \ \ \ \frho_\sigma(\feps)=\frac{1}{\Omega} \sum\limits_{p=1}^{\NipFE} \delta (\feps-\feps^p)\fsigma^p[\feps^p(t)]
  \OmegaFE^p, \label{barsig}
\end{equation}
where the notation \f{\fsigma^p[\feps^p(t)]} denotes that \f{\fsigma^p} is a functional of the whole deformation history at integration point \f{p} of the finite element model. The strain values \f{\feps^p} in Eq.~\eqref{rhodelta} and \eqref{barsig} can be interpreted as integration points in strain space. In contrast to the real space integration points~\f{\fx^p}, their strain space counterparts~\f{\feps^p=\feps(\fx^p)} are not fixed but travel in strain space during the deformation process. By replacing \f{\fsigma^p} by \f{\fsigma^p:\tcEps_k(\fx^p)} in Eq.~\eqref{barsig}, the residuals~\f{R_k} (Eq.~\eqref{discweakform}) can also be expressed as integrals in strain space. A key observation is that the exact locations \f{\fx^p} of the integration points, the grain shapes, their relative location etc.~become obsolete, because their effect on the mechanical fields is fully encoded in the strain fluctuation modes. In other words, the reduced-order problem can be fully solved in strain space and real space data is no longer required, once the strain fluctuation modes are known. \\
Like in real space, the choice of the strain space integration points is non-unique. In general, two finite element discretizations of the same microstructure will lead to different macroscopic stresses~\f{\bar \fsigma} in Eq.~\eqref{barsig}, but the difference will be negligible, if both meshes are sufficiently fine. Upon continuous mesh refinement (with arbitrarily large \f{\NipFE} and arbitrarily small \f{\OmegaFE^p}), the continuum model is in principle recovered, with continuous densities \f{\rho(\feps)} and \f{\frho_\sigma(\feps)}. Two models with different meshes will be called {\it macroscopically indistinguishable} in the following, if their macroscopic stress responses are equal, independent of the macroscopic strain history.\\
One may go even one step further and consider macroscopically indistinguishable microstructures. These define, for example, a macroscopically homogeneous solid. That is, two RVEs, taken from two arbitrary macroscopic locations, always exhibit equal effective stress responses despite their dissimilar microstructures.\\
These reflections illustrate that the \f{\NipFE} strain space integration points \f{\feps^p} in Eqns.~\eqref{rhodelta} and \eqref{barsig} are not unique. Under certain conditions, they may be replaced by other integration points
\begin{equation}
 \rho(\feps)= \frac{1}{\Omega} \sum\limits_{p=1}^{\NipFE} \delta (\feps^p-\feps) \OmegaFE^p \ \ \ 
 \rightarrow \ \ \  \rho(\feps)= \frac{1}{\Omega} \sum\limits_{q=1}^{\NipHR} \delta (\feps^q-\feps) \Omega^q
\end{equation}
without changing the macroscopic response. The new \f{\NipHR} integration points do not need to be a subset of the original \f{\NipFE} points, which are anyway non-unique. This implies that the new integration points will, in general, no longer be identified with any location~\f{\fx} in real space.\\
A naturally arising question is how many integration points (in strain space) are necessary to obtain a reduced model, which is macroscopically indistinguishable from the original finite element model.
To answer this question, it is helpful to note that the trajectories of the fully integrated reduced order model's integration points in strain space are encoded in terms of the strain fluctuation modes \f{\tcEps_k(\fx^p)}:
\begin{equation}
 \feps^p=\bar \feps + \sum_{k=1}^{\Nmd}\xi_k(t) \tcEps_k (\fx^p).\label{epspmodes}
\end{equation}
In other words, the search for a minimum set of strain space integration points essentially amounts to the question of how to replace the \f{\tcEps_k(\fx^p)} (\f{p=1,\dots,\NipFE}) by corresponding values \f{\tcEps_k^q} (\f{q=1,\dots,\NipHR}) with \f{\NipHR\ll\NipFE}, while keeping the the model macroscopically indistinguishable from (or at least close to) the finite element model. In addition, each integration point is also equipped with a certain crystal orientation~\f{\fQ^q}. If less integration points than grains are sought for, then the latter need to be identified, in addition.\\
Finally, it would be desirable to identify a  set of reduced strain fluctuation modes / strain space integration points, which do not only lead to the same macroscopic response as the finite element model, but also produce the same mode coefficients \f{\xi_k} as the fully integrated ROM, in order to enable a fast reconstruction of the microscopic field quantities.\\
\subsubsection{Identification of the E3C-modes and reduced crystal orientations}\label{sectident}
The aim in the following will be to choose the E3C-modes \f{\tcEps_k} (\f{k=1,\dots,\Nmd}), collected in the matrix \f{\tcEps} (Eq.~\eqref{tildeeps}), and grain orientations \f{\fQ^q} (\f{q=1,\dots,\NipHR}) in such a way as to identify a reduced-order model that approximates (available data of) the original finite element model as good as possible. 
To this end, \f{\Nsim} representative simulations are to be performed with {\it i}\,) the finite element model, {\it ii}\,) the fully integrated ROM and {\it iii}\,) the E3C-model. After step {\it i}\,), finite element snapshots are available, which enable to directly construct the fully integrated ROM. In addition, the FEM simulations deliver macroscopic stress values \f{\bar \fsigma^{{\rm FE},s}_{n+1}}, where \f{s\in\{1,\dots,\Nsim\}} is the simulation index and \f{n+1\in\{1,\dots,N_t^s\}} designates the time step. In step {\it ii}\,), the same macroscopic strains~\f{\bar \feps^s_{n+1}} as in step {\it i}\,) are applied to the fully integrated ROM (see Eq.~\eqref{discweakform}), which delivers for each time step \f{n\to n+1} of every simulation \f{s} the microscopic equilibrium state, encoded in the equilibrium mode coefficients \f{\fxi^s_{n+1}}.\\
Based on these data, the \f{\Nsim} simulations can be repreated with the E3C-model, prescribing the strains at the E3C integration points as
\begin{equation}
 \feps^{qs}_{n+1} = \bar \feps^s_{n+1} + \sum_{k=1}^{\Nmd}\xi^s_{k(n+1)} \tcEps^q_k \ \  \ (q=1,\dots,\NipHR).
\end{equation}
In general, the stresses \f{\fsigma^{qs}_{n+1}}, obtained by this procedure, will not be in equilibrium, i.e., Eq.~\eqref{apprs}\f{_1} is not satisfied, since the \f{\xi^s_{k(n+1)}} do not correspond to the equilibrium mode coefficients, but are taken from the fully integrated ROM solution. However, one can expect the stress fluctuations \f{\tfsigma^{qs}_{n+1}:=\fsigma^{qs}_{n+1}-\bar \fsigma^{{\rm FE},s}_{n+1}} to reside in a low dimensional subspace, since the strain fluctuations exhibit the very same property. In other words, the singular value decomposition of the matrix 
\begin{equation}
 \tilde \fs (\tcEps_k,\fQ^q) = \begin{pmatrix}
               \tfsigma^{11}_1 & \tfsigma^{11}_2 & \dots \tfsigma^{1\Nsim}_{N_t^s} \\
               \tfsigma^{21}_1 & \tfsigma^{21}_2 & \dots \tfsigma^{2\Nsim}_{N_t^s} \\
                \vdots & \vdots & \ddots & \vdots \\
                \tfsigma^{\NipHR 1}_1 & \tfsigma^{\NipHR 1}_2 & \dots \tfsigma^{\NipHR \Nsim}_{N_t^s}                
              \end{pmatrix}=\tilde\fS\fSigma\T{\fV}
\end{equation}
can be truncated after a (hopefully small) number \f{\Nmd^\sigma \leq 6 \NipHR} of singular values\footnote{Assuming \f{\tilde \fs} to be a wide matrix with more columns than rows.}. Consequently, the stress fluctuations can be represented as a linear combination of the first \f{\Nmd^\sigma} left singular values \f{\tilde \fS=(\tilde \fS_1,\dots,\tilde \fS_{\Nmd^\sigma})} of \f{\tilde \fs}:
\begin{equation}
 \begin{pmatrix}
    \tfsigma^{1}  \\
    \tfsigma^{2}  \\
    \vdots \\
    \tfsigma^{\NipHR} 
 \end{pmatrix}
 =\sum\limits_{l=1}^{\Nmd^\sigma} \mu_l\tilde \fS_l.
\end{equation}
The E3C-method aims at determining the strain fluctuation modes \f{\tcEps_k} and grain orientations~\f{\fQ^q}, such that the reduced stress and strain modes are orthogonal in the sense:
\begin{equation}
 \tilde \fS_l \cdot \fOmega \fEps_k=0, \ k=1,\dots,\Nmd+6, \ l=1,\dots,\Nmd^\sigma,\label{orthogonalitycond}
\end{equation}
where \f{\fEps_k} are the strain fluctuation and macroscopic strain modes, collected columnwise in the matrix \f{\fEps=(\tcEps \bar \fEps)} (see Eq.~\eqref{Edecomp}), and \f{\fOmega} is the diagonal matrix of the integration volumes~\f{\Omega^q}. 
By multiplying Eq.~\eqref{orthogonalitycond} with the stress fluctuation mode coefficients~\f{\mu_l}, summing over~\f{l}, and exploiting Eq.~\eqref{epscontst} one arrives at the conclusion, that 
\begin{equation}
 R_k = \frac{1}{\Omega}\sum\limits_{q=1}^{\NipHR} \tcH_k^q  :  \fsigma^q \, \Omega^q=0,  \ \ \ \bar \fsigma^{\rm FE} = \frac{1}{\Omega} \sum\limits_{q=1}^{\NipHR} \fsigma^q \, \Omega^q.
\end{equation}
{\it Provided the orthogonality condition~\eqref{orthogonalitycond} holds, it is thus ensured that a) the equilibrium mode coefficients \f{\fxi} of the fully integrated ROM and of the E3C model are equal and b) the macroscopic stresses predicted by the E3C-method coincide with those of the finite element model.}\vspace{2mm}\\
Indeed, by multiplying the orthogonality condition \eqref{orthogonalitycond} by the singular values \f{\Sigma_l}, squaring and summing over both indices, it follows that
\begin{align}
 &\tilde \fS_l \cdot \fOmega \fEps_k=0 \ \  \Leftrightarrow \ \ \sum\limits_{k=1}^{\Nmd+6} \sum\limits_{l=1}^{\Nmd^\sigma} \left( \fEps_k \cdot  \fOmega \Sigma_l\tilde \fS_l \right)^2 
 = \sum\limits_{k=1}^{\Nmd+6} \fEps_k \cdot \fOmega\tilde \fS \fSigma \underbrace{\T{\fV}\fV}_{\blds I} \T{\fSigma} \T{ \tilde \fS}\fOmega \fEps_k \label{costfctnderiv1} \\
 = &\sum\limits_{k=1}^{\Nmd} \tcEps_k \cdot \fOmega \tilde \fs \T{\tilde \fs} \fOmega\tcEps_k + \sum\limits_{k=1}^{6} \bar \fEps_k \cdot  \fOmega\tilde \fs \T{\tilde \fs} \fOmega \bar \fEps_k  \\
 = &\Omega^2 \sum\limits_{s=1}^{\Nsim} \sum\limits_{n=0}^{N_t^s-1} \left[ \sum_{k=1}^{\Nmd} \left( \frac{1}{\Omega} \sum\limits_{q=1}^{\NipHR} \fsigma^{qs}_{n+1} : \tcEps^q_k \, \Omega^q\right)^2 + \left\| \bar \fsigma^{{\rm E3C},s}_{n+1} - \bar \fsigma^{{\rm FE},s}_{n+1} \right\|^2\right]=0. \label{costfctnderiv3}
\end{align}
This very last result again illustrates that the orthogonality condition \eqref{orthogonalitycond} ensures that a) the modes~\f{\fxi^s_{n+1}} of the fully integrated ROM and the E3C model coincide and b) the E3C macro-stress equals the FEM macro-stress, provided all but \f{\Nmd^\sigma} singular values of \f{\tilde \fs} are negligible. If the truncated singular values are finite, Eq.~\eqref{orthogonalitycond} still holds approximately.\\
In order to identify strain space integration points (in terms of the \f{(\tcEps_k,\fQ^q)}), which fulfill the orthogonality condition~\eqref{orthogonalitycond}, the equivalence in Eq.~\eqref{costfctnderiv1} motivates the minimization of the left hand side of Eq.~\eqref{costfctnderiv3} (to bring it as close to the right hand side as possible). The resulting cost function is composed of two contributions:
\begin{enumerate}
 \item {\bf Macroscopic stress error.} This cost function contribution measures the distance in the macroscopic stress prediction as obtained by the finite element model and the E3C model for some given values of the E3C-modes \f{\tcEps_k} and crystal orientations \f{\fQ^q}:
 \begin{equation}
  \csig (\tcEps^q_k,\fQ^q)= \frac{1}{2} \sum\limits_{s=1}^{\Nsim} \sum\limits_{n=0}^{N_t^s-1} \left\| \underbrace{\bar \fsigma^{{\rm E3C},s}_{n+1} - \bar \fsigma^{{\rm FE},s}_{n+1}}_{\Delta \bar {\blds \sigma}^s_{n+1}} \right\|^2. \label{csig}
 \end{equation}
 The finite element stresses \f{\bar \fsigma^{{\rm FE},s}_{n+1}} are given from step {\it i}\,), while their E3C counterparts are computed as follows:
 \begin{equation}
  \bar \fsigma^{{\rm E3C},s}_{n+1} = \frac{1}{\Omega}\sum\limits_{q=1}^{\NipHR} \fsigma(\feps^{qs}_{n+1},\fX^{qs}_n,\fQ^q) \, \Omega^q 
  \ \ \ {\rm with} \ \feps^{qs}_{n+1}=\bar \feps^s_{n+1} + \underbrace{\sum\limits_{k=1}^{\Nmd}\xi^s_{k(n+1)}\tcEps^q_k}_{\tilde {\blds \eps}^{qs}_{n+1}}. \label{sigE3C}
 \end{equation}
 It is noted that the evaluation of Eq.~\eqref{sigE3C}\f{_1} requires the application of a crystal plasticity time integration algorithm, which will be outlined in Sect.~\ref{timeint}, to repeat all \f{\Nsim} simulations. In contrast, Eq.~\eqref{sigE3C}\f{_2} can be directly evaluated, as the mode coefficients~\f{\xi^s_{k(n+1)}} are known from step {\it ii}\,).
 \item{\bf Micro-equilibrium error.} The second cost function contribution is concerned with the error in the micro-state prediction of the E3C model. Here, the 'micro-state' is encoded in the vector of mode coefficients \f{\fxi}, which is computed as solution of the equilibrium condition \f{\fR(\fxi)=\fzero} (see the discussion after Eq.~\eqref{resid}). In an ideal situation, the mode coefficients predicted by the fully integrated ROM and the E3C model would coincide. However, the E3C model will usually not be that accurate, such that the E3C-residual vector will in general not vanish, if it is evaluated with the solution \f{\fxi^s_{n+1}} obtained from the fully integrated ROM. This error describes the distance of the E3C model from equilibrium and is measured by the second contribution to the cost function:
 \begin{equation}
  \cres (\tcEps^q_k,\fQ^q)= \frac{1}{2} \sum\limits_{s=1}^{\Nsim} \sum\limits_{n=0}^{N_t^s-1} \|\fR^{{\rm E3C},s}_{n+1} \|^2, 
  \ \ \ R^{{\rm E3C},s}_{k(n+1)} = \frac{1}{\Omega} \sum\limits_{q=1}^{\NipHR} \fsigma^q(\feps^{qs}_{n+1},\fX^{qs}_n,\fQ^q) : \tcEps^q_k \, \Omega^q. \label{cR}
 \end{equation}
\end{enumerate}
The overall cost function is obtained as \f{c=\csig+\cres}, yielding the minimization problem
\begin{equation}
 \{\tcEps^q_k, \fQ^q\} = {\rm arg} \ \mini{ {\tilde {\blds {\sty \cE}}^q_k} \atop {\avg{\tilde{\blds{\sty \cE}}_k}={\blds 0}} } \ \mini{{\blds Q}^q \in Orth^+} c(\tcEps^q_k, \fQ^q),\label{miniproblem}
\end{equation}
where \f{Orth^+} is the set of proper orthogonal tensors.\\
In the following section, the Levenberg Marquardt algorithm, as one possible cost function minization approach, is adopted to the minization problem~\eqref{miniproblem}. Readers not interested in the intricate technical details may skip this section and move directly to the results in Sect.~\ref{sect2scale}. However, the derivation also includes several interesting novel aspects. Amongst other things, the usual algorithmic tangent is generalized and the mechanical response is not only linearized with respect to the current state, but with respect to the whole deformation history.\\
{\bf Remark.} In case of monotonuous loading, the information stored in the full-field simulation data of a given simulation \f{s} (with \f{N_t^s} time steps) is to some extend redundant, in particular if small time steps are used. In this case, it is computationally more efficient to include only a subset of the \f{N_t^s} time steps in the cost function.
\section{Cost function minimization by the Levenberg Marquardt algorithm}
The minization of the cost function via the Levenberg Marquardt algorithm requires a time discretization and a non-standard linearization of the crystal plasticity equations, which are set forth in a preparatory step in Sect.~\ref{numcp}. The actual Levenberg Marquardt algorithm is described thereafter.
\subsection{Numerical treatment of the crystal plasticity model}\label{numcp}
\subsubsection{Time integration}\label{timeint}
The constitutive equations in Box~2 are discretized in time by the implicit Euler method. That is, all equations in the box are evaluated at time \f{t_{n+1}} and the rates are approximated as
\begin{equation}
 \dot \gamma_\alpha \approx \frac{\Delta \gamma_\alpha}{\Delta t} = \sign{\fsigma_{n+1} : \Ms} \dot \gamma_0 \left\langle \frac{|\fsigma_{n+1} : \Ms|-\tauc_{n+1}}{\tauD} \right\rangle^p , \ \ \ \Delta \gammaacc=\sum\limits_{\alpha=1}^{\Nslp} |\Delta \gamma_\alpha| \label{rates}
\end{equation}
with \f{\Delta \gamma_\alpha = \gamma_{\alpha,n+1}-\gamma_{\alpha,n}}. The vector of primary unknowns, given by~\f{\fS_{n+1}=(\fsigma_{n+1},\tauc_{n+1})}, is the solution of the nonlinear equation set \f{\fR(\fS_{n+1})=(\fr^\sigma,r^{\tauc})=\fzero} with (see Eqns.~\eqref{Hookeslawandshearstresses} and~\eqref{hardeninglaw})
\begin{align}
 \fr^\sigma &=-\ffC:\big(\feps_{n+1}-\fepsp(\fsigma_{n+1},\tauc_{n+1})\big)+\fsigma_{n+1} = \fzero, \label{rsigma}\\
 r^{\tauc} &= \ttauc \big(\gammaacc(\fsigma_{n+1},\tauc_{n+1})\big)-\tauc_{n+1} = 0, \label{rtauc}
\end{align}
that is solved via the Newton scheme in combination with a simple line search algorithm.
In Eqns.~\eqref{rsigma} and \eqref{rtauc}, the plastic strain \f{\fepsp(\fsigma_{n+1},\tauc_{n+1})=\fepsp_n+\Delta \fepsp(\fsigma_{n+1},\tauc_{n+1})} and accumulated plastic slip \f{\gammaacc(\fsigma_{n+1},\tauc_{n+1})=\gammaacc_n+\Delta \gammaacc(\fsigma_{n+1},\tauc_{n+1})} are computed according to Eqns.~\eqref{kinematics} and \eqref{rates}.
\subsubsection{Linearization}\label{lineariz}
The algorithm outlined in Sect.~\ref{timeint} delivers the stresses \f{\fS_{n+1}=(\fsigma_{n+1},\tauc_{n+1})\in \ffR^7} and internal variables \f{\fX_{n+1}=(\fepsp_{n+1},\gammaacc_{n+1})} as functions of the (input) variables \f{\fZ=(\feps_{n+1}, \fepsp_n, \gammaacc_n)}.
The assembly of the global stiffness matrix requires the algorithmic tangent \f{\ffC^{\rm a}=\d \fsigma_{n+1}/\d \feps_{n+1}}. As will be discussed below, the empirically corrected cluster cubature (E3C) method also requires additional algorithmic tangents (during the offline stage), as derived in the following.\\
Requiring the linearised residual \f{\fR} to vanish, leads to 
\begin{equation}
 \Delta \fR = \pd{\fR}{\fS_{n+1}} \Delta \fS_{n+1} + \pd{\fR}{\fZ}\Delta \fZ =\fzero \ \ \ \Rightarrow 
 \pd{\fS_{n+1}}{\fZ}=-\left( \pd{\fR}{\fS_{n+1}} \right)^{-1} \pd{\fR}{\fZ} \label{deltaR}
\end{equation}
with the involved matrices being given explicitly in \ref{appalgotang}. Note that the aforementioned algorithmic tangent~\f{\ffC^{\rm a}} emerges as the upper left 6\f{\times}6-submatrix of \f{\partial \fS_{n+1}/\partial \fZ}. The linearization of the internal variables can be expressed as
\begin{equation}
 \Delta \fX_{n+1} = \Delta \fX_n + \pd{\fX_{n+1}}{\fS_{n+1}} \pd{\fS_{n+1}}{\fZ} \Delta \fZ \ \ \ 
 \Rightarrow \pd{\fX_{n+1}}{\fZ}=(\fzero_{7\times6}, \ \fI_{7\times7}) + \pd{\fX_{n+1}}{\fS_{n+1}} \pd{\fS_{n+1}}{\fZ},
\end{equation}
where the entries of the matrix
\begin{equation}
 \pd{\fX_{n+1}}{\fS_{n+1}} = \begin{pmatrix}
                   \pd{\fepsp_{n+1}}{\fsigma_{n+1}} & \pd{\fepsp_{n+1}}{\tauc_{n+1}} \vspace{2mm} \\
                   \pd{\gammaacc_{n+1}}{\fsigma_{n+1}} & \pd{\gammaacc_{n+1}}{\tauc_{n+1}} 
                 \end{pmatrix}
\end{equation}
are also given in \ref{appalgotang}.
\subsubsection{Linearization with respect to the crystal orientation}\label{linori}
Besides the input parameters \f{\fZ}, the E3C-method also interprets the crystal orientation \f{\fQ} as input parameter and requires a related linearization:
\begin{equation}
 \Delta \fQ = \hat \fomega_{\Delta} \fQ,
\end{equation}
where \f{\hat \fomega_\Delta = -\fepsilon \fomega_\Delta} is skew-symmetric (due to the skew-symmetry of \f{\Delta \fQ\T{\fQ}}) and the vector~\f{\fomega_\Delta} describes the incremental rotation angle \f{\|\fomega_\Delta\|} and rotation axis \f{\fomega_\Delta/\|\fomega_\Delta\|}. Based on these notions, linearizations of the output stresses \f{\fS_{n+1}} and internal variables \f{\fX_{n+1}} can be specified:
\begin{equation}
 \Delta \fS_{n+1} = D_{\omega} \fS_{n+1} \, \fomega_\Delta, \ \ \ \Delta \fX_{n+1} = {D_\omega} \fX_{n+1} \, \fomega_\Delta,
\end{equation}
where the matrices \f{D_{\omega} \fS_{n+1} \in \ffR^{7\times 3}} and \f{D_\omega \fX_{n+1} \in \ffR^{7\times 3}} are derived in \ref{applino}.
\subsection{General Levenberg Marquardt algorithm}\label{lmalgo}
The minimization problem \eqref{miniproblem} is a nonlinear least-squares parameter fitting problem. Many well established solution methods exist for this problem class, ranging from direct search methods over gradient methods to the widespread Gauss-Newton method. A particularly robust approach is the Levenberg-Marquardt (LM) algorithm \citep{levenberg1944method,marquardt1963algorithm}, which will be applied in this work. While the classical Newton method would require a laborious computation of the the cost function's Hessian, the LM-algorithm only needs its gradient \f{\partial c(\fY)/\partial \fY}, where the parameter vector \f{\fY} is given by
\begin{equation}
 \fY=\begin{pmatrix}
      \tcEps^1\\
      \fP^1\\
      \vdots\\
      \tcEps^{\NipHR}\\
      \fP^{\NipHR}\\
     \end{pmatrix} \in \ffR^{\NipHR\cdot(6\Nmd+3)} \ \ \ {\rm with} \
 \tcEps^q=\begin{pmatrix}
           \tcEps^q_1\\
           \vdots\\
           \tcEps^q_{\Nmd}
          \end{pmatrix} \in \ffR^{6\Nmd}.
\end{equation}
The parameters \f{\fP^q\in\ffR^3} represent a general parametrization of the rotation tensors \f{\fQ^q}. Different choices are in principle possible (e.g., Euler angles, unit quaternions, ...). The choice made in this work is further explained below. By eliminating \f{\tcEps^{\NipHR}} from \f{\fY}, the reduced parameter vector \f{\hat \fY} is defined in addition (with \f{6\Nmd} parameters less), accounting for the possibility to reconstruct \f{\tcEps^{\NipHR}} at any time as a result of constraint \eqref{epscontst}:
\begin{equation}
 \tcEps^{\NipHR} = -\frac{1}{\Omega^{\NipHR}}\sum\limits_{q=1}^{\NipHR-1} \tcEps^q \Omega^q.
\end{equation}
A Levenberg-Marquardt solution step consists in solving the following linear system of equations
\begin{equation}
 \underbrace{\hat \fG}_{\partial c /\partial \hat {\blds Y}} + (\hat\fH + \lambda \fI) \Delta \hat \fY = \fzero, \label{LMstep}
\end{equation}
where \f{\hat\fH} approximates the Hessian of \f{c(\hat \fY)} and \f{\lambda} is a damping factor, such that for \f{\lambda=0}, the classical Newton method is approximately recovered. The quantities~\f{\hat \fG} and \f{\hat \fH} are given by the relations
\begin{equation}
 \hat \fG = \T{\fL} \fG, \ \ \ \hat \fH = \T{\fL} \fH \fL, \ \ \ {\rm with} \ \fL=\pd{\fY}{\hat \fY} \label{GandHhat}
\end{equation}
and 
\begin{align}
 \fG=&\pd{c}{\fY} =  \sum\limits_{s=1}^{\Nsim} \sum\limits_{n=0}^{N_t^s-1} \T{ \left( \pd{\Delta \bar \fsigma^s_{n+1}}{\fY} \right) } \Delta \bar \fsigma^s_{n+1} 
 + \T{ \left( \pd{\fR^{{\rm E3C},s}_{n+1}}{\fY} \right) } \fR^{{\rm E3C},s}_{n+1},  \label{eqG}\\
 \fH=&  \sum\limits_{s=1}^{\Nsim} \sum\limits_{n=0}^{N_t^s-1} \T{ \left( \pd{\Delta \bar \fsigma^s_{n+1}}{\fY} \right) } \pd{\Delta \bar \fsigma^s_{n+1}}{\fY} 
 + \T{ \left( \pd{\fR^{{\rm E3C},s}_{n+1}}{\fY} \right) } \pd{\fR^{{\rm E3C},s}_{n+1}}{\fY}.  \label{eqH}
\end{align}
To enable damping in~\f{\fY} rather than in~\f{\hat \fY}, the term \f{\lambda\fI} in Eq.~\eqref{LMstep} is replaced by \f{\lambda\T{\fL}\fL} in this work. The damping factor is continuously updated following the heuristics in \citet{nielsen1999damping} and \citet{gavin2019levenberg}, Sect.~4.1.1 (algorithm 3). 
\subsection{Derivative computation}
The derivatives with respect to the parameter vector \f{\fY} require the linearization of the crystal plasticity algorithm (implemented as material subroutine, see Sect.~\ref{timeint}), which provides the stress \f{\fsigma_{n+1}} and internal variables \f{\fX_{n+1}}, requires the input parameters \f{\{\feps_{n+1}, \fX_n, \fP\}} and can thus formally be written as
\begin{equation}
 \fsigma_{n+1}(\feps_{n+1}, \fX_n, \fP), \ \ \ \fX_{n+1}(\feps_{n+1}, \fX_n, \fP).
\end{equation}
During the cost function minimization process, the input parameters take the form \f{\{\feps_{n+1}, \fX_n, \fP\}} \f{=\{\feps^{qs}_{n+1}, \fX^{qs}_n, \fP^q\}} and the linearization of the stress is given by
\begin{align}
 \Delta \fsigma^{qs}_{n+1} = & \pd{\fsigma_{n+1}}{\feps_{n+1}} \Delta \feps^{qs}_{n+1}
                              +\pd{\fsigma_{n+1}}{\fX_n} \Delta \fX^{qs}_{n}
                              +\pd{\fsigma_{n+1}}{\fP} \Delta \fP^q\\
                          \stackrel{\eqref{sigE3C}}{=} & \sum\limits_{l=1}^{\Nmd} \underbrace{\left( \pd{\fsigma_{n+1}}{\feps_{n+1}} \xi^s_{l(n+1)}
                              +\pd{\fsigma_{n+1}}{\fX_n} \pd{\fX^{qs}_n}{\tcEps^q_l} \right)}_{\partial {\blds \sigma}^{qs}_{n+1} / \partial \tilde{\blds{\sty \cE}}^q_l}  \Delta \tcEps^q_l
                              +\underbrace{\left( \pd{\fsigma_{n+1}}{\fP} + \pd{\fsigma_{n+1}}{\fX_n} \pd{\fX^{qs}_n}{\fP^q} \right)}_{\partial {\blds \sigma}^{qs}_{n+1} / \partial {\blds P}^q} \Delta \fP^q.
\end{align}
Here, the function \f{\fX^{qs}_{n+1}(\tcEps^q,\fP^q):\stackrel{\eqref{sigE3C}}{=}\fX_{n+1}(\feps^{qs}_{n+1}(\tcEps^q), \fX^{qs}_n(\tcEps^q,\fP^q), \fP^q)} is recursively defined in~\f{\fX^{qs}} with \f{\fX^{qs}_0=\fzero} (initial conditions of the internal variables). In other words, the partial derivatives with respect to the parameters \f{\fY} require a linearization with respect to the whole deformation history, which manifests itself in the complexity of the above-stated expressions.\\
The partial derivatives of the function \f{\fX^{qs}_{n+1}(\tcEps^q,\fP^q)} are also recursively defined and can be obtained in complete analogy: 
\begin{align}
                  \pd{\fX^{qs}_{n+1}}{\tcEps^q_l} &= \pd{\fX_{n+1}}{\feps_{n+1}} \xi^s_{l(n+1)}
                              +\pd{\fX_{n+1}}{\fX_n} \pd{\fX^{qs}_n}{\tcEps^q_l} ,  \\
                  \pd{\fX^{qs}_{n+1}}{\fP^q} &= \pd{\fX_{n+1}}{\fP} + \pd{\fX_{n+1}}{\fX_n} \pd{\fX^{qs}_n}{\fP^q} .
\end{align}
With these partial derivatives at hand, it is now possible to linearize the errors contributing to the cost function and to identify the partial derivatives with respect to the parameters \f{\fY} in Eqns.~\eqref{eqG} and~\eqref{eqH}:
\begin{align}
  \Delta \Delta \bar \fsigma^s_{n+1}  \stackrel{\eqref{csig},\eqref{sigE3C}}{=} 
  & \sum \limits_{q=1}^{\NipHR} \sum\limits_{l=1}^{\Nmd} \underbrace{\frac{\Omega^q}{\Omega}\pd{\fsigma^{qs}_{n+1}}{\tcEps^q_l}}_{\partial \Delta \bar {\blds \sigma}^s_{n+1} / \partial \tilde{\blds{\sty \cE}}^q_l} \Delta \tcEps^q_l
  +\underbrace{\frac{\Omega^q}{\Omega}\pd{\fsigma^{qs}_{n+1}}{\fP^q}}_{\partial \Delta \bar {\blds \sigma}^s_{n+1} / \partial {\blds P}^q} \Delta \fP^q,\\
  \Delta R^{{\rm E3C},s}_{k(n+1)} \stackrel{\eqref{cR}}{=}
  & \sum\limits_{q=1}^{\NipHR} \sum\limits_{l=1}^{\Nmd} \underbrace{\frac{\Omega^q}{\Omega} \left( \fsigma^{qs}_{n+1} \delta_{kl} + \T{ \left[ \pd{\fsigma^{qs}_{n+1}}{\tcEps^q_l} \right] } \tcEps^q_k \right)}_{\partial \Delta R^{{\rm E3C},s}_{k(n+1)} / \partial \tilde{\blds{\sty \cE}}^q_l} \Delta \tcEps^q_l
  +\underbrace{\frac{\Omega^q}{\Omega} \left( \T{ \left[ \pd{\fsigma^{qs}_{n+1}}{\fP^q} \right] } \tcEps^q_k \right) }_{\partial \Delta R^{{\rm E3C},s}_{k(n+1)} / \partial {\blds P}^q} \Delta \fP^q. \label{deltaReq}
\end{align}
{\bf Remark.} As is observable from Eqns.~\eqref{GandHhat} to \eqref{deltaReq}, the algorithmic tangent \f{\partial \fsigma_{n+1}/\partial \feps_{n+1}} is needed for the computation of the residual~\f{\hat \fG}. As a result, kinks in the stress strain relation lead to discontinuities in the residual, which are expected to thwart the performance of the Levenberg Marquardt algorithm. If kinks\footnote{Sharp kinks will not occur due to the power law flow rule \eqref{flowrule}, but the effect may be similar.} do occur, for example due to other hardening laws (e.g., Voce) or due to cyclic loading, it might thus be necessary to modify the solution procedure, e.g., by excluding time steps with elastoplastic transition from the optimization process or by regularizing the hardening law. However, these cases have not been investigated, yet.
\subsection{Algorithm}\label{sectAlgo}
The overall assembly of the vector \f{\fG} and matrix \f{\fH} (Eqns.~\eqref{eqG} and~\eqref{eqH}), needed to accomplish a Levenberg-Marquardt step (see Eq.~\eqref{LMstep}), is summarized in Box~3. This algorithmic realization of Eqns.~\eqref{eqG} to \eqref{deltaReq} involves a multitude of single crystal plasticity simulations (with prescribed strain paths), which may be parallelized.\\
If the linearization of the crystal orientations is parametrized via the vectors \f{\fomega_\Delta^q} (replacing \f{\Delta \fP^q}), as described in Sect.~\ref{linori}, then the matrices \f{\partial \fsigma_{n+1}/\partial \fP} and \f{\partial \fX_{n+1}/\partial \fP} must be replaced by \f{D_\omega \fsigma_{n+1}} and \f{D_\omega \fX_{n+1}}, respectively. In this case, the update \f{\fP^q\leftarrow\fP^q+\Delta \fP^q} after a successful Levenberg Marquardt step is replaced by \f{\fQ^q \leftarrow \exp{(\hat \fomega_\Delta^q)}\fQ^q}, where the well-known Rodrigues formula is used for the matrix exponential.
\begin{figure}
\centering
\framebox{
\footnotesize
\begin{minipage}{.9\textwidth}
\vspace{3mm}
\setstretch{1.8}
 {\tt Input:} \f{\bar \fsigma^{{\rm FE},s}_{n+1}}, \f{\bar \feps^s_{n+1}}, \f{\fxi^s_{n+1}}, \f{\tcEps^q_k}, \f{\fP^q}, \f{\Omega^q}, \f{\Omega}\\
 {\tt Output:} \f{\csig}, \f{\cres}, \f{\fG}, \f{\fH}\\
 \f{\csig \leftarrow 0}; \f{\cres \leftarrow 0}; \f{\fG\leftarrow\fzero}; \f{\fH\leftarrow\fzero}\\
 {\tt for} \f{s=1,\dots,\Nsim} \{ {\tt //This loop may be parallelized.}
 \begin{adjustwidth}{15pt}{0pt}  
  \f{\fX_0\leftarrow \fzero}\\
  {\tt for} \f{n=0,\dots,N_t^s-1} \{
  \begin{adjustwidth}{15pt}{0pt}  
   \f{\fR^{{\rm E3C},s}_{n+1}\leftarrow\fzero}; \ \ \f{\pd{\fR^{{\rm E3C},s}_{n+1}}{\fY}\leftarrow\fzero}; \ \ \f{\Delta \bar \fsigma^s_{n+1} \leftarrow -\bar \fsigma^{{\rm FE},s}_{n+1}}; \ \ \f{\pd{\Delta \bar \fsigma^s_{n+1}}{\fY}\leftarrow\fzero}\\
   {\tt for} \f{q=1,\dots,\NipHR} \{
   \begin{adjustwidth}{15pt}{0pt}  
     \f{\feps^{qs}_{n+1} \leftarrow \bar \feps^s_{n+1} + \sum \limits_{k=1}^{\Nmd}\xi^s_{k(n+1)} \tcEps^q_k}\\
     {\tt call material subroutine (input:} \f{\feps^{qs}_{n+1}}, \f{\fX^{qs}_n}, \f{\fP^q} )\\
      \f{\ \ \ \rightarrow} \f{\fsigma_{n+1}}, \f{\pd{\fsigma_{n+1}}{\feps_{n+1}}}, \f{\pd{\fsigma_{n+1}}{\fX_n}}, \f{\pd{\fsigma_{n+1}}{\fP}}, \f{\pd{\fX_{n+1}}{\feps_{n+1}}}, \f{\pd{\fX_{n+1}}{\fX_n}}, \f{\pd{\fX_{n+1}}{\fP}}\\
      \f{\Delta \bar \fsigma^s_{n+1} \peq \frac{\Omega^q}{\Omega}\fsigma_{n+1}}\\
      {\tt for} \f{k=1,\dots,\Nmd+1} \{ {\tt //The last run is related to }\f{\fP^q}.
      \begin{adjustwidth}{15pt}{0pt}  
       {\tt if} \f{k\leq \Nmd} \bigg\{ \f{\pd{\fsigma^{qs}_{n+1}}{\tcEps^q_k} \leftarrow \xi^s_{k(n+1)} \pd{\fsigma_{n+1}}{\feps_{n+1}}} \bigg\} 
       {\tt else} \bigg\{ \f{\pd{\fsigma^{qs}_{n+1}}{\fP^q} \leftarrow \pd{\fsigma_{n+1}}{\fP}} \bigg\}\\
       {\tt if} \f{n=0} \bigg\{ \f{\pd{\fX^{qs}_{n+1}}{\sty \cZ}\leftarrow\fzero} \bigg\} {\tt //with} \f{{\sty \cZ}=\tcEps^q_k} {\tt or} \f{{\sty \cZ}=\fP^q} {\tt depending on }\f{k}\\
       {\tt else} \bigg\{
         \f{\pd{\fsigma^{qs}_{n+1}}{\sty \cZ} \peq \pd{\fsigma_{n+1}}{\fX_n}\pd{\fX^{qs}_n}{\sty \cZ}}; 
         \f{\pd{\fX^{qs}_{n+1}}{\sty \cZ} \leftarrow \pd{\fX_{n+1}}{\fX_n}\pd{\fX^{qs}_n}{\sty \cZ}}
         \bigg\}\\
         {\tt if} \f{k\leq \Nmd}\{
         \begin{adjustwidth}{15pt}{0pt} 
           \f{\pd{\fX^{qs}_{n+1}}{\tcEps^q_k} \peq \xi^s_{k(n+1)} \pd{\fX_{n+1}}{\feps_{n+1}}}; \ \ \
           \f{R^{{\rm E3C},s}_{k(n+1)} \peq \fsigma_{n+1} : \tcEps^q_k \frac{\Omega^q}{\Omega}}; 
           \ \ \ \f{\pd{\Delta \bar \fsigma^s_{n+1}}{\tcEps^q_k}\leftarrow\frac{\Omega^q}{\Omega} \pd{\fsigma^{qs}_{n+1}}{\tcEps^q_k}}\\
           {\tt for} \f{l=1,\dots,\Nmd} \bigg\{ 
             \f{\pd{R^{{\rm E3C},s}_{l(n+1)}}{\tcEps^q_k} \leftarrow \T{ \left( \pd{\fsigma^{qs}_{n+1}}{\tcEps^q_k} \right) } \tcEps^q_l \frac{\Omega^q}{\Omega} }
           \bigg\}\\
           \f{\pd{R^{{\rm E3C},s}_{k(n+1)}}{\tcEps^q_k} \peq \fsigma_{n+1} \frac{\Omega^q}{\Omega}} 
         \end{adjustwidth} 
         \}\\
         {\tt else} \{ 
         \begin{adjustwidth}{15pt}{0pt}
          \f{\pd{\fX^{qs}_{n+1}}{\fP^q} \peq \pd{\fX_{n+1}}{\fP}}; 
          \ \ \ \f{\pd{\Delta \bar \fsigma^s_{n+1}}{\fP^q}\leftarrow\frac{\Omega^q}{\Omega} \pd{\fsigma^{qs}_{n+1}}{\fP^q}}\\
          {\tt for} \f{l=1,\dots,\Nmd} \bigg\{ 
             \f{\pd{R^{{\rm E3C},s}_{l(n+1)}}{\fP^q} \leftarrow \T{ \left( \pd{\fsigma^{qs}_{n+1}}{\fP^q} \right) } \tcEps^q_l \frac{\Omega^q}{\Omega} }
          \bigg\}
         \end{adjustwidth}
         \}
      \end{adjustwidth}
      \vspace{-2mm}
      \}
   \end{adjustwidth}  
   \vspace{-2mm}
   \}\\
  \f{\csig \peq \frac{1}{2}\|\fR^{{\rm E3C},s}_{n+1}\|^2; \ \ \ \cres \peq \frac{1}{2}\|\Delta \bar \fsigma^{qs}_{n+1}\|^2};
  \ \ \ \f{\fG \peq \T{ \left( \pd{\Delta \bar \fsigma^s_{n+1}}{\fY} \right) } \Delta \bar \fsigma^s_{n+1} + \T{ \left( \pd{\fR^{{\rm E3C},s}_{n+1}}{\fY} \right) } \fR^{{\rm E3C},s}_{n+1} }\\
  \f{\fH \peq   \T{ \left( \pd{\Delta \bar \fsigma^s_{n+1}}{\fY} \right) } \pd{\Delta \bar \fsigma^s_{n+1}}{\fY} 
 + \T{ \left( \pd{\fR^{{\rm E3C},s}_{n+1}}{\fY} \right) } \pd{\fR^{{\rm E3C},s}_{n+1}}{\fY} }
  \end{adjustwidth}  
  \}
 \end{adjustwidth}
 \vspace{-2mm}
 \}
 \vspace{1mm}
\end{minipage}
}
\caption*{Box~3: Assembly algorithm for LM-step.}
\end{figure}\\
{\bf Starting guess.} The Levenberg Marquardt algorithm requires a starting guess for the E3C-modes \f{\tcEps_k} and crystal orientations \f{\fQ^q}. This starting solution can have a substantial influence on the quality of the result, if the cost function exhibits several minima with different cost function values. The simple strategy followed here consists in clustering grains together, such that their combined volumes \f{\Omega^p} (\f{p=1,\dots,\NipHR}) are approximately equal. This strategy is chosen to 'eliminate' small grains with potentially minor influence on the effective behavior. A greedy algorithm is used, which starts with empty clusters and always puts the next largest grain into the (up to that point) smallest cluster, starting with the largest grain. Subsequently, all grains in a given cluster \f{\cC^q} are merged by averaging their crystal orientations \f{\fQ_i} according to the minization problem \citep{markley2007averaging}
\begin{equation}
 \fQ^q = {\rm arg} \mini{{\blds \fQ}\in {Orth}^+} \sum \limits_{i\in\cC^q} \| \fQ_i - \fQ \|^2 \Omega_i,
\end{equation}
where lower indices mark the original (unclustered) grains and upper indices refer to their clustered counterparts.
The granular volumes are not included in the list of cost function parameters, because additional constraints (\f{\Omega^q\geq0}) would then potentially further complicate the optimization process.\\
The starting guess for the \f{\tcEps^q_k} is obtained by cluster-wise averaging the \f{\tcEps_k(\fx)} of the fully integrated ROM. After this clustering process, the cost function minimization modifies the modes \f{\tcEps_k} and orientations \f{\fQ^q} according to the data collected by the finite element model and the fully integrated ROM, hence the name 'Empirically Corrected Cluster Cubature (E3C)'.
\section{Two-scale simulation procedure}
Once the E3C modes and grain orientations are available, a two-scale simulation can be performed in an FE\f{^2}-like fashion. In order to solve the macroscopic problem numerically via the Newton method, the E3C model is called at each macroscopic Gauss point in each Newton iteration. The required data comprise the (current) macroscopic strain \f{\bar \feps_{n+1}} and the mode coefficients \f{\fxi_n} (as starting guess for the local Newton scheme), the stresses~\f{\fsigma^q_n} and~\f{\tau^{{\rm c},q}_n} (as starting guess for the Newton scheme on the grain level) and internal variables \f{\fX_n^q} (\f{q=1,\dots,\NipHR}). At each macroscopic Gauss point, a local Newton method determines the mode coefficients \f{\fxi_{n+1}}. The related residual vector \f{\fR(\fxi)} and Jacobian \f{\fK=\partial \fR/\partial \fxi} are assembled by looping over all \f{\NipHR} E3C-grains and calling the related crystal plasticity time integration algorithm (see Sect.~\ref{numcp}) with strains \f{\feps^q_{n+1}} computed according to Eq.~\eqref{microstrain}. That crystal plasticity algorithm is called \f{\NipHR} times at each macroscopic integration point and delivers the quantities required for the assembly of the residual and the Jacobian:
\begin{equation}
 R_k = \frac{1}{\Omega} \sum\limits_{q=1}^{\NipHR} \fsigma^q_{n+1}:\tcEps^q_k \Omega^q; \  \ \ 
 K_{kl} = \frac{1}{\Omega} \sum\limits_{q=1}^{\NipHR} \tcEps^q_k: \left(\frac{\d \fsigma_{n+1}}{\d \feps_{n+1}}\right)^q : \tcEps^q_l \Omega^q.
\end{equation}
The macroscopic algorithmic tangent is obtained from the linearization of the macro-stress:
\begin{equation}
 \Delta \bar \fsigma_{n+1} = \underbrace{\frac{1}{\Omega} \sum\limits_{q=1}^{\NipHR} \left(\frac{\d \fsigma_{n+1}}{\d \feps_{n+1}}\right)^q : \left(\ffI^{\rm s} +  \sum\limits_{k=1}^{\Nmd} \tcEps_k^q \otimes \frac{\d \xi_{k,n+1}}{\d \bar \feps_{n+1}} \right) \Omega^q}_{\d \bar {\blds \sigma}_{n+1}/\d \bar {\blds \eps}_{n+1}}: \Delta \bar \feps_{n+1}.\label{macrotang}
\end{equation}
The final required derivative \f{\d \xi_{k,n+1}/\d \bar \feps_{n+1}} is obtained from the linearization of the residual~\f{R_k}:
\begin{align}
 \Delta R_k &= \pd{R_k}{\bar \feps_{n+1}}:\Delta \bar \feps_{n+1} + \sum\limits_{l=1}^{\Nmd} K_{kl} \Delta \xi_{l,n+1}  =0\\
 \Leftrightarrow \frac{\d \xi_{k,n+1}}{\d \bar \feps_{n+1}} &= \frac{1}{\Omega} \sum\limits_{l=1}^{\Nmd} (K^{-1})_{kl}  \sum\limits_{q=1}^{\NipHR} \T{\left(\frac{\d \fsigma_{n+1}}{\d \feps_{n+1}}\right)^q} \!\!\!\!\! : \tcEps^q_l \Omega^q.
\end{align}
Symmetry of the macroscopic algorithmic tangent may be shown upon insertion of this result into Eq.~\eqref{macrotang} (the crystal plasticity tangent \f{\d \fsigma_{n+1}/\d \feps_{n+1}} is symmetric, too).

\section{Two-scale simulations}\label{sect2scale}
\subsection{Plate with a hole}
The first example is a plate with a hole. This academic study aims to provide a simple introduction of the overall procedure in order to make the reader familiar with the general workflow. To this end, several simplifications are made. A more complex and practically more relevant example is provided in Sect.~\ref{sectMP}.
\subsubsection{Macroscale simulation setup}
\begin{figure}[h] 
\centering
  \psfrag{x}{\f{x}}
  \psfrag{y}{\f{y}}
  \psfrag{z}{\f{z}}
\hspace{-10mm}\ifthenelse{\boolean{badQuality}}{
  \includegraphics[width=0.9\textwidth]{Figs/LochplatteVollSmall}
}{
  \includegraphics[width=0.9\textwidth]{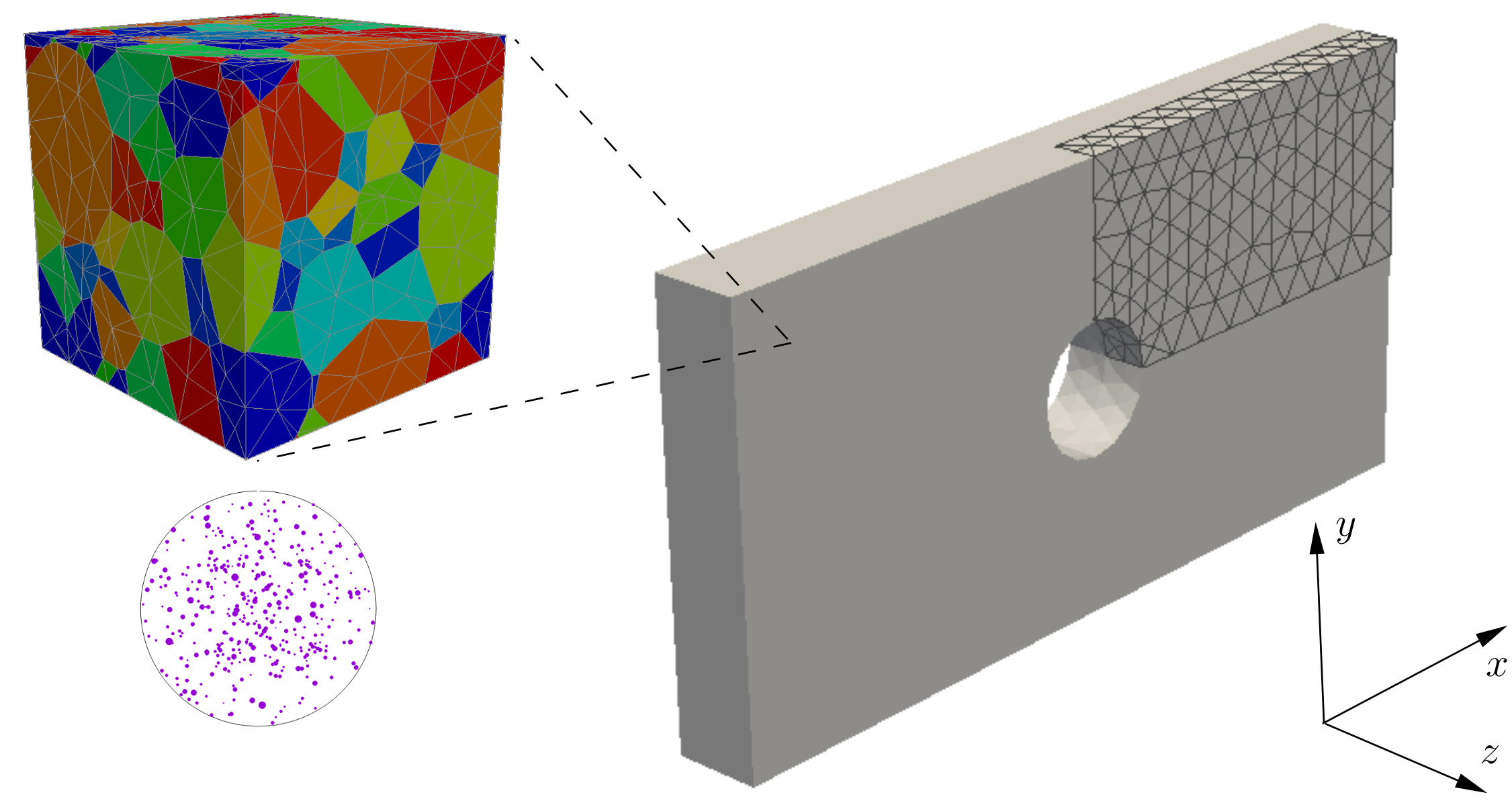}
}
\caption{Plate with a hole with microstructure, consisting of 99 grains. The meshed region of the plate has dimensions 10\f{\times}5\f{\times}1~mm\f{^3}. The \{111\} pole figure (bottom left) illustrates a nearly isotropic crystal orientation distribution, where the dots are scaled according to the grain sizes.}
\label{LochplatteVoll}
\end{figure}

Only one quarter of the plate is simulated, for simplicity (see Fig.~\ref{LochplatteVoll}, right), using symmetry boundary conditions. The displacement in \f{x}-direction is prescribed at the rear end, reaching a maximum value of 0.1~mm after 1~s and allowing for lateral contractions.
\\
For simplicity, linear tetrahedra are used on the macroscale despite their relatively poor performance.
\subsubsection{Microscale simulation setup}
The microscopic model is also depicted in Fig.~\ref{LochplatteVoll} (left) and consists of 99 periodic face centered cubic grains with 12 slip systems each, which are meshed by quadratic tetrahedra, leading to a total of 35643 displacement degrees of freedom\footnote{For the geometry generation, the software Neper (see neper.info/papers.html) served as a basis. The meshes were generated using gmsh 4.11.1 \citep{geuzaine2009gmsh}.}. The corresponding \{111\} pole figure is also depicted and illustrates an almost isotropic orientation distribution. The single crystal material parameters are summarized in Tab.~\ref{matparam}.
\begin{table}[h]
  \centering
  \begin{tabular}{||c|c|c|c|c|c|c|c||}
  \hline
    \f{E} [GPa] & \f{\nu} [-] & \f{\dot \gamma_0} [s\f{^{-1}}] & \f{\tauD} [MPa] & \f{p} [-] & \f{q} [MPa] & \f{n} [-] & \f{\gamma_0} [-] \\ \hline
    70 & 0.35 & 0.001 & 50 & 20 & 70 & 0.2 & 0.001\\
  \hline
  \end{tabular}
  \caption{Material parameters of the crystal plasticity model summarized in Box~2, where \f{E} and \f{\nu} denote Young's modulus and Poisson's ratio.}
  \label{matparam}
\end{table}
\subsubsection{Training -- Mode identification, E3C hyper-reduction and computational costs}
\begin{figure}[h]
\centering
  \includegraphics[width=0.95\textwidth]{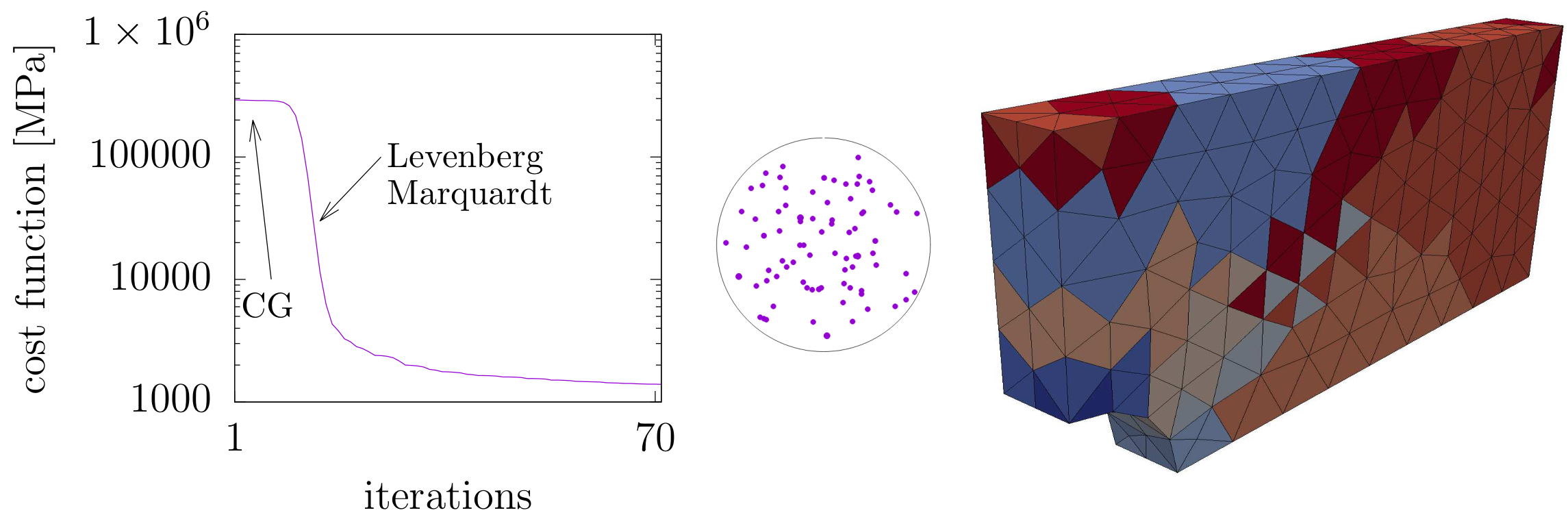}
\caption{Left: Cost function evolution for 20 modes and 20 E3C integration points. The first few iterations are carried out by the conjugate gradient method ('CG'), which was observed to show slower convergence than the subsequently used Levenberg Marquardt algorithm. The maximum iteration number was set to 70. Center: \{111\} pole figure illustrating the 20 grain orientations obtained by the E3C cost function minimization. Right: Clusters obtained by k-means for retraining.}
\label{costfctn20}
\end{figure}

The training of the reduced order model requires finite element simulation data. Here, 31 microscopic simulations with prescribed macrostrain \f{\bar \feps(t)=(t/t_{\rm max})\cdot \bar \feps_0} with \f{t_{\rm max}=1}~s and strain magnitude \f{\|\bar \feps_0\|=0.05} were carried out. The 31 values for \f{\bar \feps_0} were identified by a modified latin hypercube sampling procedure (see \ref{applhc}). All simulations were conducted using adaptive time steps with an initial and maximum time step of \f{0.05}~s, leading to a total number of 20 time steps for most simulations, as convergence was usually achieved on the first try. The typical wall clock time\footnote{The CPU used throughout this work was an Intel$^{\textregistered}$ Core\f{^{\rm TM}} i7-8850H CPU @ 2.60GHz with 32 GB RAM (standard laptop hardware). The FEM simulations were done using FEAP 8.5 \citep{taylor2014feap}.} of a single simulation is 8~min, summing up to 31\f{\times}8~min\f{\approx}4.2~h for the snapshot collection. The nodal displacement fluctuations were collected in a snapshot matrix, which was subsequently further processed by proper orthogonal decomposition (CPU time: \f{\sim}1~min) in order to identify the displacement/strain modes, forming the 'fully integrated' reduced order model (ROM). Using 20 modes, the 31 strain paths mentioned above were re-simulated using this ROM (total CPU time: \f{\sim}20~min) in order to collect the mode coefficients \f{\xi^s_{k(n+1)}} (\f{s=1,\dots,31}, \f{n=0,\dots,N^s_t-1}, \f{k=1,\dots,20}). The latter supplement the related macroscopic stresses (from the finite element model\footnote{In the rare cases, where the finite element model required more than 20 time steps, the corresponding macro-stresses from the ROM (which always converged on the first try) were used.}) in order to provide the data required for the subsequent computation of the E3C modes via minimization of the cost function (see Sect.~\ref{sectident}), using the Levenberg Marquardt algorithm. For 20 E3C integration points (i.e., the 99 originial grains are merged into 20 grains as described in Sect.~\ref{sectAlgo}), the cost function minimization took \f{\sim}9~min and the corresponding cost function evolution is depicted in the semi-logarithmic diagram in Fig.~\ref{costfctn20} (left). As can be observed, the Levenberg Marquardt algorithm results in a quick decrease of the cost function, followed by a rather flat section of the curve. Experience shows that the online performance of the E3C modes does not significantly change during this last phase. Out of the twenty time steps, the cost function included time steps 5,10,15 and 20. The center of the figure depicts the \{111\} pole figure resulting from the E3C cost function minimization, indicating no observable texture of the almost equally weighted 20 grains.
\subsubsection{Clustered retraining}
\begin{figure}[h] 
\centering
\includegraphics[width=0.99\textwidth]{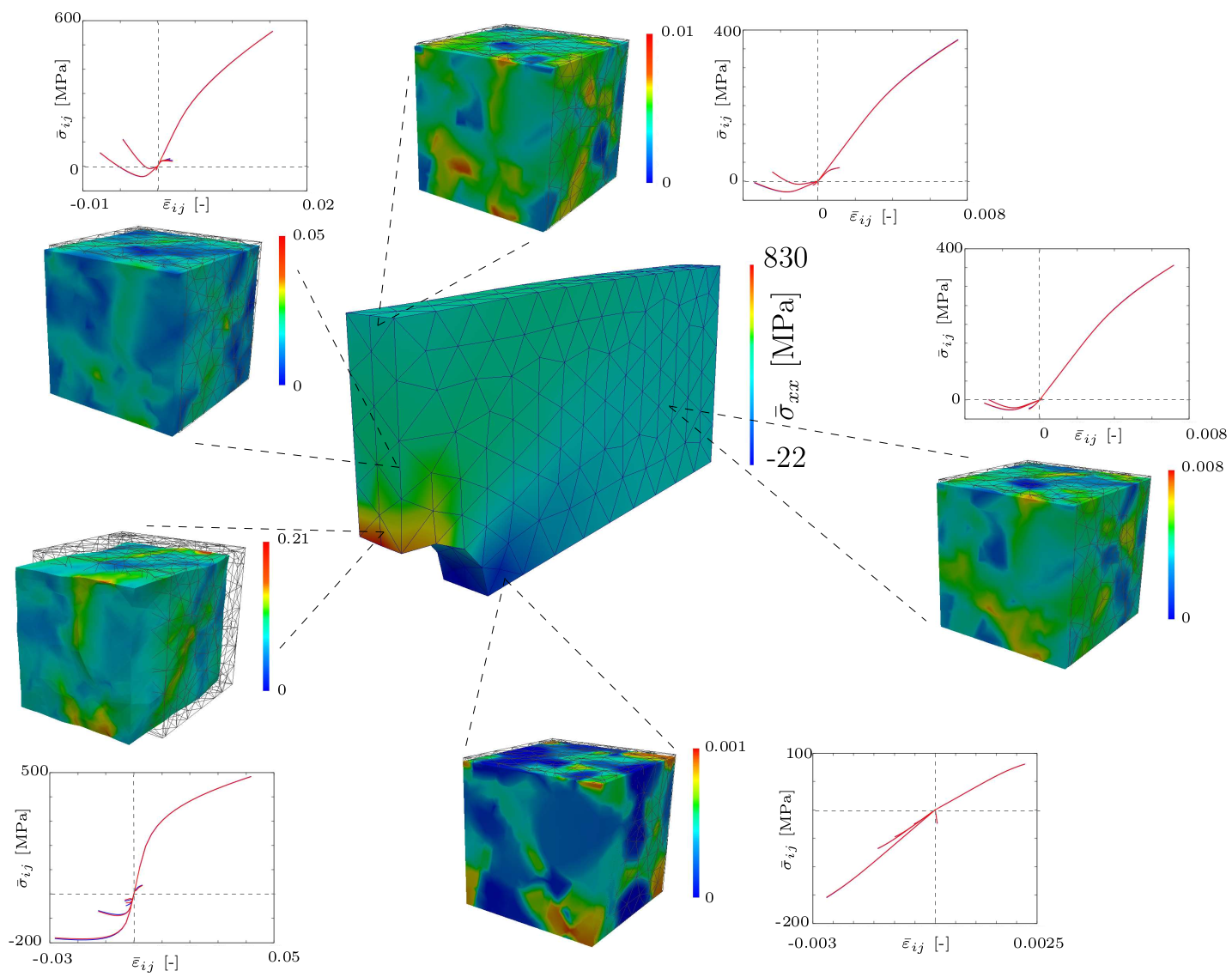}
\caption{Final state of the 'plate with a hole' two-scale simulation (deformations scaled by factor 5) with an online wall-clock time of \f{\sim}2.5~min (without reconstruction). The results were obtained after clustered training and show the microscopic deformations (FEM reconstructions showing the accumulated plastic slip~\f{\gammaacc}) and corresponding macroscopic stress-strain curves (in MPa) for several macro elements. The red curves show the E3C-predictions, while the corresponding FEM-results, indicated in blue, are hardly visible due to the good match with the E3C-results. The \f{\sigma_{xx}/\eps_{xx}}-curves (tension) are those in the first quadrant, while the lateral responses (contraction) can be found in the third quadrant. The shear components are small in magnitude, but are also included in the diagrams for completeness.}
\label{overviewpic}
\end{figure}

Once, the E3C modes are available, a two-scale simulation can be run. To this end, in each of the 655 macroscopic elements, an individual microscopic problem is solved concurrently with the macroscopic problem in an FE\f{^2}-like fashion. The primary microscopic variables are the mode coefficients~\f{\xi_k} (here: \f{k=1,...,20}), and the assembly of the corresponding residuals \f{R_k} requires the crystal plasticity material law evaluation in all 20 E3C integration points. Figure \ref{overviewpic} shows an overview of the macroscopic results and five exemplary microstates at the end of a similar simulation, which will be explained in detail in Sect.~\ref{SimRes} below.\\
Once the two-scale problem is solved, one may either accept the solution or try to further improve the accuracy by exploiting the information obtained from the two-scale simulation in order to retrain the microscopic problem and adapt it specifically to the macro-problem at hand. To this end, the snapshots (from the first 31 FEM simulations) are extended by carrying out further microscopic finite element simulations with prescribed macro-strains taken from the two-scale simulation. With these additional snapshots, the data base for the ROM identification is extended and adapted to the specific two-scale problem to be solved. Here, 10 additional microscopic FEM simulations are carried out with prescribed macroscopic strains~\f{\bar \feps_0} obtained from the (preliminary) two-scale simulation by a clustering approach, similar to the one proposed by \citet{lange2024monolithic} (see also \citet{kalina2023fe}). For this purpose, the macroscopic strain tensors at the end of the two-scale simulation are clustered by the k-means algorithm \citep{macqueen1967some} (i.e., the clustering is carried out in the six-dimensional strain space). The ten obtained clusters are illustrated in Fig.~\ref{costfctn20} (right), effectively grouping elements together, which exhibit similar strains. The average final strain of each cluster (i.e., the cluster center) is then used as the desired final macroscopic strain \f{\bar \feps_0} for the additional 10 FEM simulations (wall clock time: \f{\sim}1.3~h) to be carried out for the snapshot matrix extension.
\\
To capture the additional information in the resulting extended snapshot matrix, 30 modes and 30 integration points are chosen for the retrained E3C model, leading to additional offline costs, essentially comprising \f{\sim}42~min for the fully integrated ROM simulations and \f{\sim}56~min for the cost function minimization (both wall clock time). The main contributions to the computational effort are summarized in Tab.~\ref{traintimes}. It can be observed that the finite element computations represent the largest part (\f{\gtrsim}70\%). In summary, the total training effort is in the order of 4-5~h without retraining and 6-7~h with retraining. In comparison, the online effort (wall clock time: \f{\sim}2.5~min) for the two-scale simulation is negligible in this example. Significant potential for reducing the offline wall clock time exists. In particular, the FEM simulations -- currently the main bottleneck -- were carried out sequentially (one simulation after the other), and are expected to be straightforwardly parallelizable on more advanced desktop hardware.
\begin{table}[h]
  \centering
  \begin{tabular}{||c||c|c|c||}
  \hline
   & FEM & Full ROM & empir.~correction  \\ \hline \hline
   before retraining & 4.2~h (31 sim.) & 20~min (31 sim., 20 modes) & 9~min (20 modes, 20 IPs) \\ \hline
   retraining & 1.3~h (10 sim.) & 42~min (41 sim., 30 modes) & 56~min (30 modes, 30 IPs) \\
  \hline
  \end{tabular}
  \caption{Computational effort (approx.~wall clock time). 'FEM': Snapshot collection via FEM; 'Full ROM': fully intergrated ROM simulations (same macro-strain paths as FEM); 'empir.~correction': cost function minimization via the Levenberg Marquardt algorithm. Negligible contributions to the computational effort (e.g., POD/mode computation, k-means clustering, online effort) are not listed.}
  \label{traintimes}
\end{table}
\subsubsection{Simulation Results}\label{SimRes}
\begin{figure}[h]
\centering
  \includegraphics[width=0.99\textwidth]{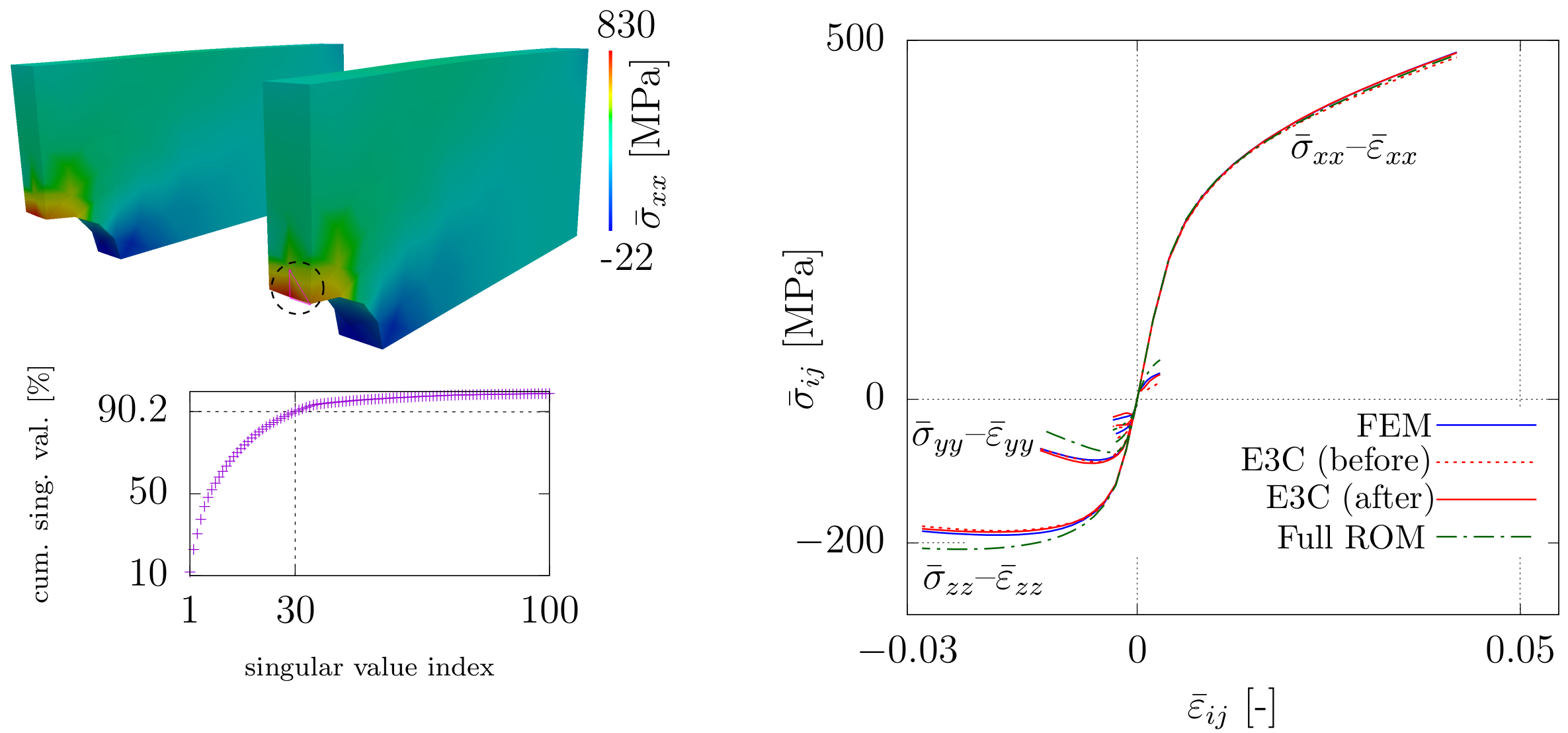}
\caption{Top left: Final macrostate before (front) and after clustered training (back). Right: Corresponding stress strain curves for the element marked in the left picture. Bottom left: First 100 cumulated singular values of the snapshot matrix, indicating that a high accuracy of the fully integrated ROM can be expected to require an increased number of modes.}
\label{cmprsnBeforeAfterClstrdTrnng}
\end{figure}

The final macroscopic state at maximum elongation is illustrated in Fig.~\ref{overviewpic} (center). The results were obtained using five time steps, the macroscopic (relative) residuals of which are given in Tab~\ref{resids}. After the two-scale simulation, the microscopic solutions for individual macro-elements can be reconstructed in different ways. For simplicity, the reconstruction is done by taking the final macro-strain for a given element and assuming that this strain state is approximately reached as a linear function of time. In a post-processing step, a corresponding microscopic FEM simulation can then be carried out. The five micro-simulations depicted in Fig.~\ref{overviewpic} were carried out in this manner. The figure also contains the related macroscopic stress-strain diagrams for both, the finite element model (blue curves) and the E3C model (red curves). The curves are hardly differentiable. Only for the lower left case, a minor discrepancy is observable.
\begin{table}[h]
  \centering
  \begin{tabular}{||c|c|c|c|c||}
  \hline
1.00E+00 & 1.00E+00 & 1.00E+00 & 1.00E+00 & 1.00E+00     \\
1.06E-01 & 7.96E-03 & 7.32E-03 & 3.85E-03 & 2.86E-03     \\
6.52E-02 & 1.04E-04 & 4.32E-05 & 1.71E-05 & 1.50E-05      \\
3.36E-02 & 6.90E-08 & 7.19E-09 & 1.66E-09 & 3.20E-09        \\
5.64E-03 & 4.37E-14 &                &                &          \\
1.67E-05 &                &                &                &           \\
1.29E-09 &                &                &                &            \\
  \hline
  \end{tabular}
  \caption{Macroscopic relative residuals.}
  \label{resids}
\end{table}
In order to investigate the effect of the clustered retraining, the two-scale simulation results before and after retraining are compared in Fig.~\ref{cmprsnBeforeAfterClstrdTrnng}. The macro-stresses, depicted in the upper left, show that the retraining step does not lead to a notable change in the macroscale results for this example. This is confirmed by the diagram in the right part of the figure. As can be seen, the E3C model is already close to the FEM-predictions before retraining. The latter slightly improves the match with the finite element model, though.\\
In addition, the results obtained with the significantly slower fully integrated ROM (using 30 modes) are also depicted in green. The fully integrated ROM shows a larger mismatch with the FEM results than the E3C model. This might be surprising at first glance, since the E3C hyper-reduction may be understood as a further model reduction step, being derived from the fully integrated ROM. However, this image is not completely correct, as the E3C modes are trained (via cost function minimization) such as to match the macroscopic FEM stresses rather than those of the fully integrated ROM. In this sense, the E3C model seems to recover information that was lost when the fully integrated ROM was derived from the snapshots. These findings are supported by the relative cumulated singular values\footnote{The relative cumulated singular value with index \f{i} is defined as the sum of the largest \f{i} singular values divided by the sum of all singular values.} of the snapshot matrix depicted in the lower left part of the figure. The graph may be interpreted such that the first 30 modes only cover \f{\sim}~90.2~\% of the information stored in the snapshot matrix, which agrees with the observation that the fully integrated ROM with 30 modes reproduces the FEM results less accurately than the E3C model.
\subsection{Mounting plate}\label{sectMP}
\subsubsection{Simulation setup}
\begin{figure}[h] 
\centering
\includegraphics[width=0.99\textwidth]{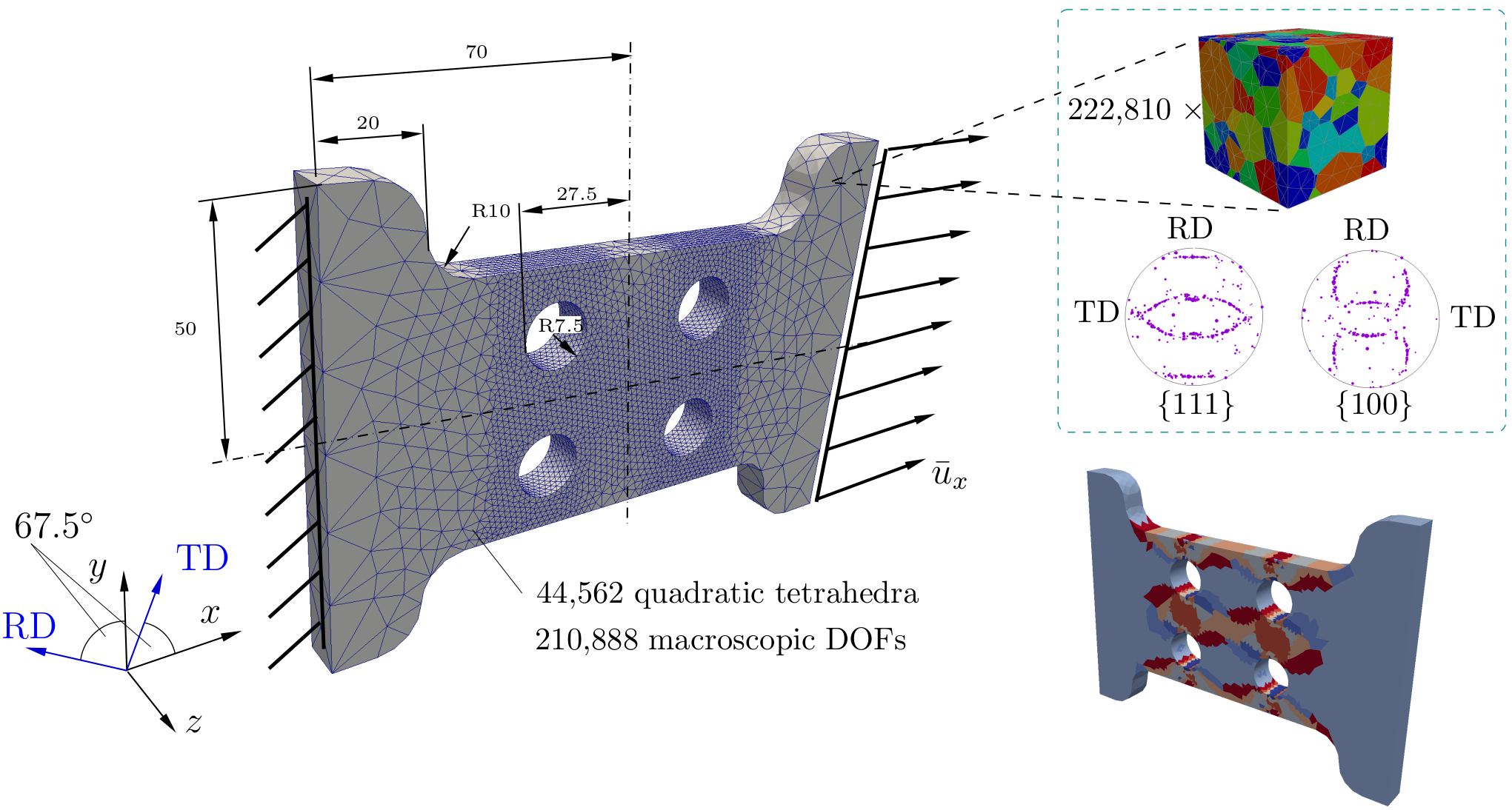}
\caption{Left: Dimensions, mesh and boundary conditions of mounting plate. Right: microstructure with \{111\} and \{100\} pole figures (top) and clusters (bottom).}
\label{masse4hole}
\end{figure}

The next example is the mounting plate depicted in Fig.~\ref{masse4hole}. For simplicity, the same 99 grains as in the previous example constitute the micro-model. However, the grain orientations are assigned such as to mimic a rolling texture, see the pole figures in Fig.~\ref{masse4hole} \citep[see, e.g., ][for comparable experimental pole figures]{hu1952rolling}. The rolling and transverse texture directions (RD/TD) are assumed to make an angle of 67.5\f{^\circ} with the mounting plate's coordinate system, breaking the symmetry of the problem (see the blue coordinate system in the figure). Symmetry boundary conditions are therefore not applied. The boundary conditions are chosen such that an overall elongation of 1.4~mm (1\% of the total lengt) is reached after 1~s, allowing for free lateral contractions, with a maximum time step of 0.05~s.
\\
The mounting plate was meshed by 44,562 quadratic tetrahedra, leading to a total number of 210,888 macroscopic displacement degrees of freedom (DOFs). Using five integration points per macro-element, the assembly of the macroscopic stiffness matrix and residual requires a total number of 5\f{\times}44,562=222,810 micro-problems to be solved via the E3C method. Since the laptop-hardware utilised had six cores, the five micro-simulations required on the element level were parallelized using OpenMP \citep{dagum1998openmp}. 
\subsubsection{Intitial training}
The E3C model was trained in complete analogy to the previous example (plate with a hole). That is, 20 modes were computed via POD from the snapshots of 31 FEM simulations (now with the textured microstructure). This results in a fully integrated ROM, which was used to repeat the 31 simulations in order to collect the equilibrium mode coefficients, which are required for the E3C mode identification via cost function minimization (see Sect.~\ref{sectident}). In a first step, 20 grains / E3C integration points were defined.
\subsubsection{Computational effort}
While the offline times are comparable to the ones mentioned for the plate with a hole (4-5~h), the online wallclock time for the two-scale simulation is now significantly higher (\f{\sim}3.7~h compared to \f{\sim}3~min for the plate with a hole), mainly due to the larger number of macroscopic integration points. For comparison, a purely macroscopic simulation (using a von Mises plasticity model), with comparable time discretization, requires approximately half an hour on the same hardware. In other words, there is a factor of approximately 7-8 between the two-scale and the single-scale simulation. 
\begin{figure}[h] 
\centering
\includegraphics[width=0.99\textwidth]{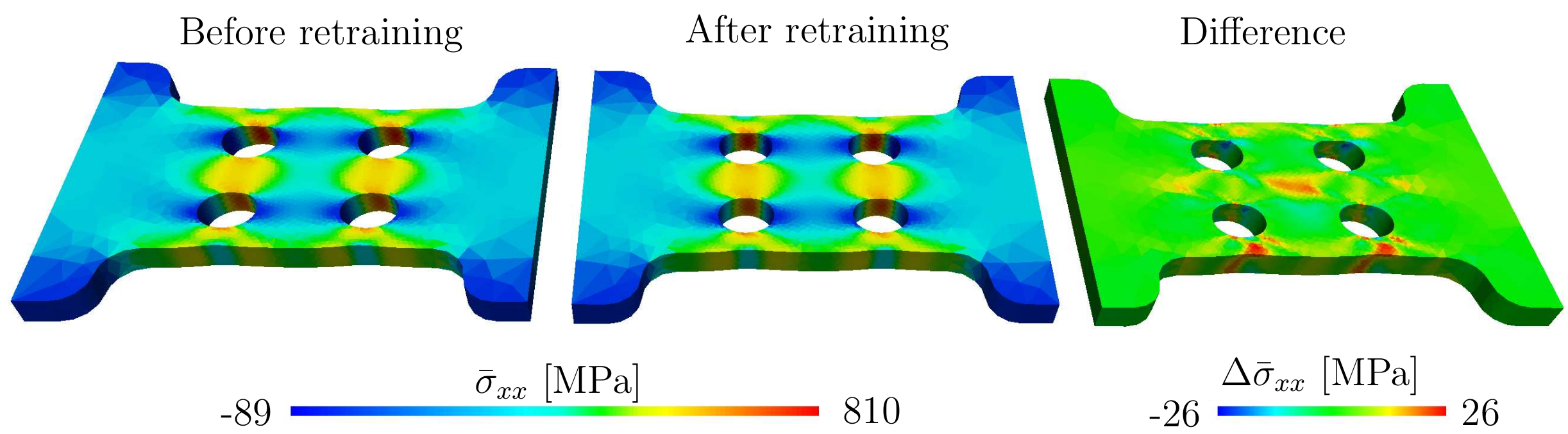}
\caption{Comparison before and after retraining (deformation scaled by factor 5).}
\label{comparisonBefAndAftRetrMountPlateSxx}
\end{figure}
\subsubsection{Clustered retraining}
The same clustered retraining strategy as in the previous example was applied, again choosing 10~clusters on the macroscale, which are visualized in Fig.~\ref{masse4hole} (bottom, right). In a first trial, the number of 20 E3C integration points / grains was kept, but led to errors of several percent in the macroscopic stress response, when being exemplarily compared to the microscopic finite element model. Therefore, the number of E3C integration points was increased to 30, while keeping a total number of 20 E3C modes, leading to an increased wall clock time of \f{\sim}5.5~h for the two-scale simulation. This corresponds to \f{\sim}148.6\% of the two-scale simulation time with 20 E3C integration points (\f{\sim}3.7~h).
\begin{figure}[h] 
\centering
\includegraphics[width=0.99\textwidth]{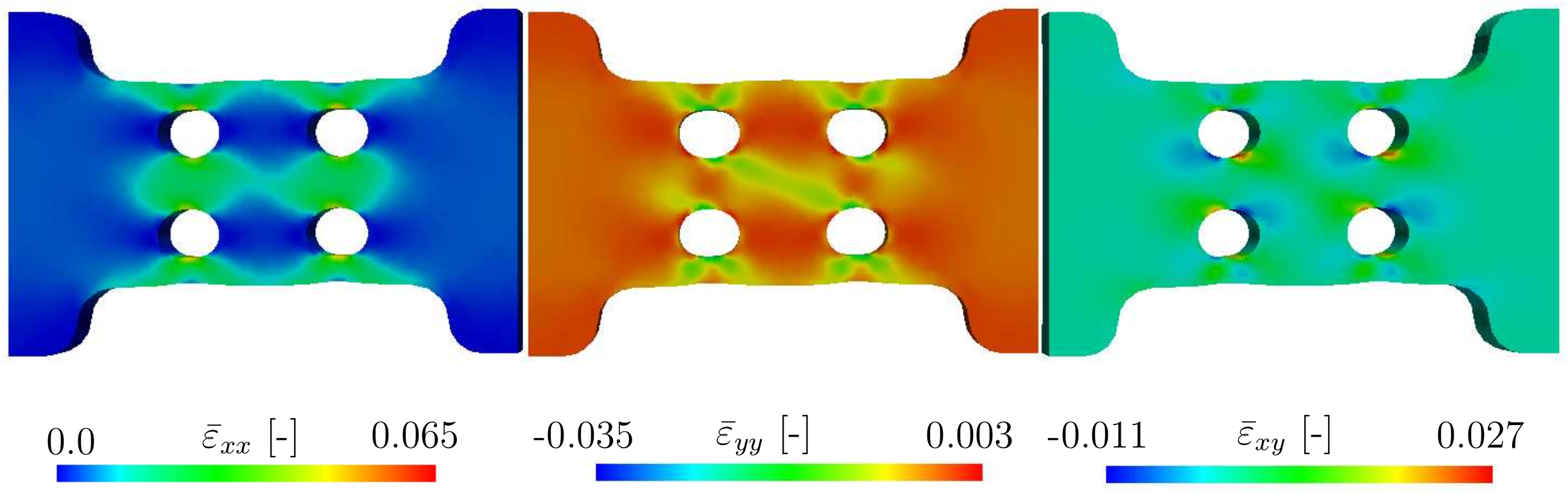}
\caption{Exemplary strain components, illustrating the influence of the texture on the deformation (deformation scaled by factor 5).}
\label{mountingPlateStrains}
\end{figure}
\subsubsection{Simulation results}
Figure \ref{comparisonBefAndAftRetrMountPlateSxx} shows a comparison of the results before (left) and after (center) clustered retraining. Since the results are hardly differentiable by the eye, the difference is also depicted (right). This indicates that the clustered retraining step may be unnecessary in many applications, depending on the accuracy requirements.\\
As mentioned before, the textured microstructure breaks the symmetry of the macroscale problem (compared to an isotropic material), which is also reflected in the strain fields depicted in Fig.~\ref{mountingPlateStrains}. In addition, it may be seen in the figure, that the displacement at the right end exhibits a small shift in the \f{y}-direction (i.e., upwards), which would not be expected in case of an isotropic material. This is also reflected in the non-symmetric shear strain distribution (rightmost in the figure) with a shift towards positive values.
\begin{figure}[h] 
\centering
\includegraphics[width=0.99\textwidth]{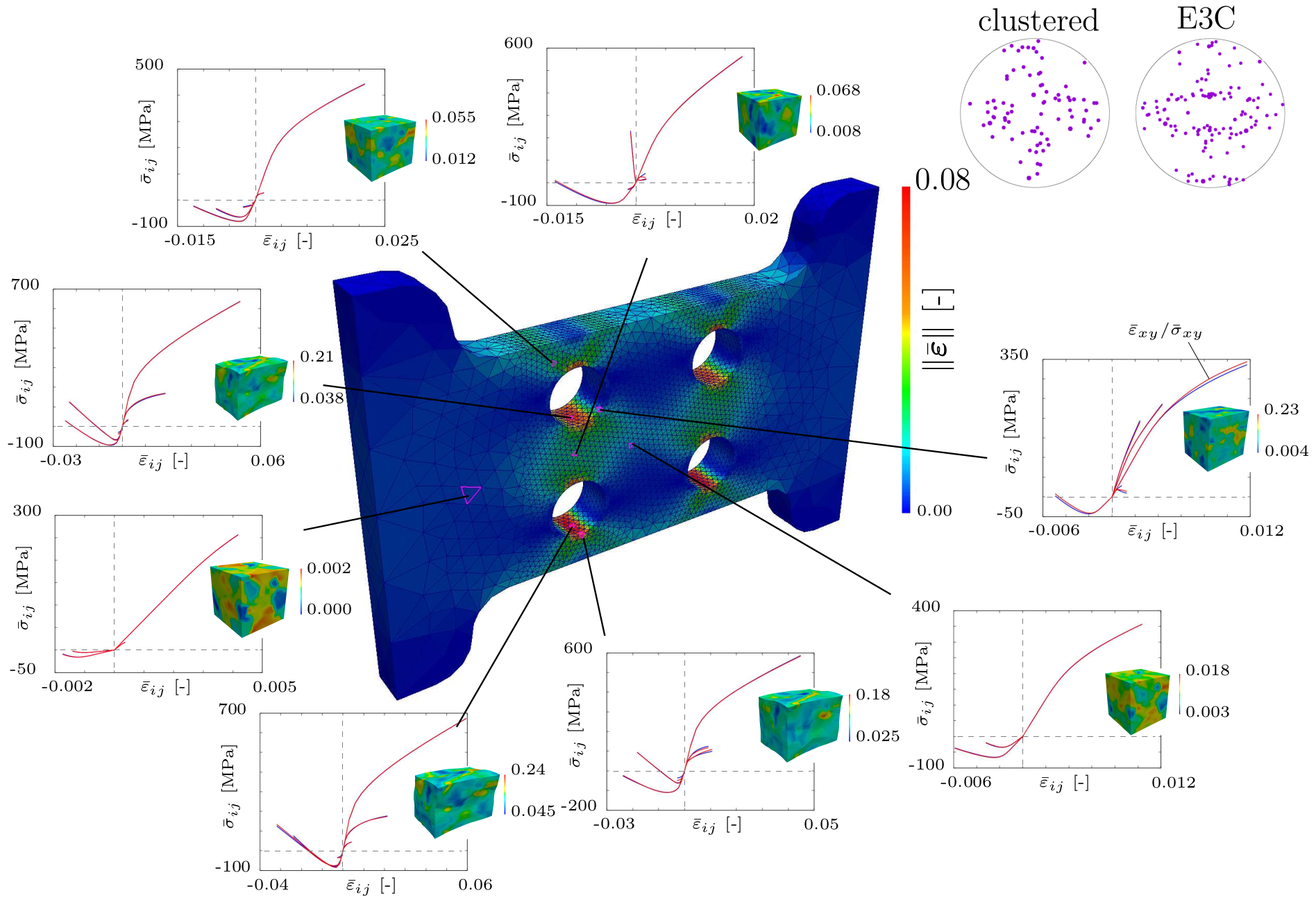}
\caption{Final state of the 'mounting plate' two-scale simulation (deformations scaled by factor 5). The results were obtained after clustered training and show the microscopic deformations (reconstructions showing the accumulated plastic slip~\f{\gammaacc}) and corresponding macroscopic stress-strain curves for several macro elements. The red curves show the E3C-predictions, while the corresponding FEM-results are indicated in blue. The \f{\sigma_{xx}/\eps_{xx}}-curves (tension) are those in the first quadrant, while the lateral responses (contraction) can be found in the third quadrant. The shear components are mostly small in magnitude, but are also included in the diagrams for completeness.}
\label{overviewpicMount}
\end{figure}\\
The microscopic fields were reconstructed in analogy to the plate with a hole, and exemplary results are depicted in Fig.~\ref{overviewpicMount} for eight positions of the mounting plate. Again, most macroscopic stress strain diagrams of the finite element model (blue curves) are hardly distinguishable from the predictions of the E3C model (red curves) despite the fact that, in some cases, the macroscopic strain norm clearly exceeds the maximum value of 0.05 seen during training. The largest error occurs for the element with the rightmost stress-strain diagram. This element is located at the edge of a hole under approximately 45\f{^\circ} with respect to the loading direction. Amongst all depicted stress-strain curves, it is the only one, where the shear component \f{\bar \eps_{xy}} dominates. In other words, the strain in this element is quite different from the majority of the strain field, which is dominated by tension in \f{x}-direction, and which the E3C model is mostly trained on during the retraining phase. This might explain the increased error at this location, which is in the order of \f{\sim 3}\% and thus similar to the deviations between the E3C-models before and after retraining (see Fig.~\ref{comparisonBefAndAftRetrMountPlateSxx}).\\
Figure \ref{overviewpicMount} also contains the \{111\} pole figure of the E3C model (top, right) in comparison with the pole figure resulting from merging the 99 grains into a smaller set of 30 grains (label 'clustered' in the figure). Obviously, the grain merging process leads to a significant loss of texture information (compare with the pole figure in Fig.~\ref{masse4hole}). However, the E3C cost function minimization seems to recover this information at least partially, as the 'E3C' texture is closer to the original one with 99 grains in Fig.~\ref{masse4hole} (right).
\subsection{Discussion}
The mounting plate example illustrates that the E3C method enables three-dimensional elastoplastic two-scale simulations of textured engineering parts with an increased number of macroscopic degrees of freedom on laptop hardware with reasonable computational effort. The online and offline efforts are well equilibriated. The main offline effort consists in the snapshot generation via microscopic FEM simulations, while the bottleneck of the online simulation is still given in terms of the micro-problems solved at the macroscopic integration points. At the same time, both contributions should be rather easily parallelizable on comparably narrow desktop hardware, indicating a high potential for further wall clock time reduction. It is expected that the two-scale simulation time will become even more competitive with the purely macroscopic single-scale computation (using von Mises plasticity). Currently, the von Mises model is still approximately one order of magnitude faster (online time) than the two-scale simulation for the discussed example. Generally, further parallelization may improve the competitiveness of the two-scale simulation, since the bottleneck of the single-scale simulation, i.e., the sparse linear equation solution, is expected to scale less beneficially with increasing parallelization than the micro-scale simulations. Further research is necessary to challenge this hypothesis.\\
The clustered retraining procedure was shown to increase the accuracy. In particular, a better agreement between reconstructed E3C and FEM results was found in terms of locally evaluated macroscopic stress strain curves (Fig.~\ref{overviewpicMount}) after retraining. However, from an engineering perspective, the overall results before and after clustered retraining (see Fig.~\ref{comparisonBefAndAftRetrMountPlateSxx}) are hard to distinguish by the eye, such that a (costly) clustered retraining step may be unnecessary for many applications.
\section{Summary}
The E3C-hyper-reduction method has been extended to crystal plasticity in this work. It is the first application of the E3C-method to history-dependent materials and the first three-dimensional mechanical implementation to be published. The method relies on reduced strain fluctuation modes, which replace the modes of the fully integrated reduced order model. A crucial observation made is that the modes lift the fully integrated ROM from real space to strain space: the shape of the grains, their relative positions and interactions are fully encoded in the modes, which renders the real space superfluous for the remaining solution process.\\
One possible interpretation of the E3C method is to view the final model as a reduced set of grains, with grain-wise constant fields. The E3C grains do not actually have a shape (but a volume) and interact through the E3C modes. In an alternative interpretation, the E3C grains are rather considered as generalized integration points in strain space. The collective motion of these integration points in strain space is defined in terms of the E3C modes and is controlled by the mode coefficients. Thus, the E3C integration points may be considered as being representatives of the integration points of the fully integrated ROM, which form a much denser point cloud in strain space, the motion of which is likewise defined in terms of the ROM's modes.\\
The E3C modes and crystal orientations are determined such as to match the FEM/ROM as accurately as possible, aiming at E3C stress and strain modes, which are orthogonal for the same states \f{\{\bar \feps,\fxi\}}. This has been found to be equivalent to the minimization of the error in the macroscopic stress and residuals for a representative set of loading scenarios. It seems that the fit to the macroscopic stresses obtained by FEM -- and {\it not} the ROM -- partially reverses inaccuracies introduced by the ROM. At least, the E3C model has been exemplarily found to be more accurate than the ROM, but more research is necessary to confirm this finding. The minimization is achieved via the Levenberg Marquardt algorithm, the adoption of which to the problem at hand has been described in detail.\\
Two scale-bridging simulations have been performed, which demonstrate that the method can handle three-dimensional engineering parts with reasonable numerical resolution on both scales, using laptop hardware. For the snapshot collection, \f{\sim}30-40 microscopic finite element simulations have been performed and have been identified as the bottleneck of the two-scale simulation. The macroscopic stress response has been evaluated at several macroscopic integration points and has been found to accurately match corresponding microscopic finite element simulations. In comparison with a single-scale simulation using von Mises plasticity, a difference in online effort of approximately one order of magnitude has been observed for the considered example. It is expected, that this factor can be decreased through further parallelization on more advanced hardware. In particular, the parallelization of the micro-simulations are assumed to be significantly more effective than that of the linear equation solver.
\begin{appendix}
\section{Algorithmic derivatives}\label{appalgotang}
The tangent required for the local Newton scheme and the linearization of the residual with respect to the input parameters \f{\fZ=(\feps_{n+1},\fepsp_n,\gammaacc_n)} are given by
\begin{equation}
 \pd{\fR}{\fS_{n+1}} = \begin{pmatrix}
                  \ffC:\pd{\fepsp_{n+1}}{\fsigma_{n+1}} + \ffI & -\ffC:\pd{\gammaacc_{n+1}}{\fsigma_{n+1}}\vspace{3mm}\\
                  \pd{\ttauc}{\gammaacc_{n+1}}\pd{\gammaacc_{n+1}}{\fsigma_{n+1}} & \pd{\ttauc}{\gammaacc_{n+1}}\pd{\gammaacc_{n+1}}{\tauc_{n+1}}-1
                 \end{pmatrix}, \ \ \
\pd{\fR}{\fZ} = \begin{pmatrix}
                  -\ffC & \ffC & \fzero\vspace{3mm}\\
                  \fzero & \fzero & \pd{\ttauc}{\gammaacc_{n+1}}
                 \end{pmatrix}                 \nonumber
\end{equation}
with
\begin{align}
 \pd{\fepsp_{n+1}}{\fsigma_{n+1}} &= \sum\limits_{\alpha=1}^{\Nslp} \pd{\gamma_{\alpha,n+1}}{\tau_{\alpha,n+1}} \Ms \otimes \Ms, \nonumber \\ 
 \pd{\gammaacc_{n+1}}{\fsigma_{n+1}} &= -\pd{\fepsp_{n+1}}{\tauc_{n+1}} =\sum\limits_{\alpha=1}^{\Nslp} \sign{\tau_{\alpha,n+1}} \pd{\gamma_{\alpha,n+1}}{\tau_{\alpha,n+1}} \Ms , \label{dXdS}\\
 \pd{\gammaacc_{n+1}}{\tauc_{n+1}} &= -\sum\limits_{\alpha=1}^{\Nslp}  \pd{\gamma_{\alpha,n+1}}{\tau_{\alpha,n+1}}. \nonumber
\end{align}
\section{Linearization with respect to the crystal orientation}\label{applino}
An increment \f{\Delta \fQ} of the crystal orientation implies corresponding increments of the slip directions and slip plane normals:
\begin{align}
 \Delta \fd_\alpha &= \Delta \fQ \fd_{\alpha 0} = \Delta \fQ \T{\fQ} \fd_\alpha = \hat \fomega_\Delta \fd_\alpha = \fomega_\Delta \times \fd_\alpha 
 = (\fomega_\Delta \cdot \fn_\alpha) \fl_\alpha - (\fomega_\Delta \cdot \fl_\alpha) \fn_\alpha \\
 &= (\fl_\alpha \otimes \fn_\alpha - \fn_\alpha \otimes \fl_\alpha ) \fomega_\Delta\\
 \Delta \fn_\alpha &= ( \fd_\alpha \otimes \fl_\alpha - \fl_\alpha \otimes \fd_\alpha ) \fomega_\Delta
\end{align}
with \f{\fl_\alpha = \fn_\alpha \times \fd_\alpha}. The linearization of the symmetric part of the Schmid tensor is given by
\begin{align}
 \Delta \Ms &= \Delta (\fsym{\fd_\alpha\otimes \fn_\alpha}) \\
 &= \underbrace{ \left[\fsym{\fl_\alpha\otimes \fn_\alpha} \otimes \fn_\alpha - \fn_\alpha\otimes \fn_\alpha \otimes \fl_\alpha + \fd_\alpha\otimes \fd_\alpha \otimes \fl_\alpha - \fsym{\fd_\alpha\otimes \fl_\alpha} \otimes \fd_\alpha  
 \right]}_{=:D_\omega {\blds M}^{\rm s}_\alpha} \fomega_\Delta.
\end{align}
The linearizations of the plastic strain and accumulated plastic slip increment are given by (see also Eqns.~\eqref{rates} and \eqref{dXdS})
{\footnotesize
\begin{align}
 \Delta \Delta \fepsp &= \pd{\fepsp_{n+1}}{\fsigma_{n+1}} \Delta \fsigma_{n+1} + \pd{\fepsp_{n+1}}{\tauc_{n+1}} \Delta \tauc_{n+1} 
                       + \underbrace{ \left( \sum\limits_{\alpha=1}^{\Nslp} \pd{\gamma_{\alpha,n+1}}{\tau_{\alpha,n+1}} \Ms \otimes (\fsigma_{n+1}:D_\omega \Ms) + \Delta \gamma_{\alpha,n+1} D_\omega \Ms \right) 
                       }_{=:D_\omega {\blds \eps}^{\rm p}} \fomega_\Delta, \nonumber\\
\Delta \Delta \gammaacc & = \pd{\gammaacc_{n+1}}{\fsigma_{n+1}} \Delta \fsigma_{n+1} + \pd{\gammaacc_{n+1}}{\tauc_{n+1}} \Delta \tauc_{n+1} 
                      + \underbrace{ \left( \sum\limits_{\alpha=1}^{\Nslp} \sign{\tau_{\alpha,n+1}} \pd{\gamma_{\alpha,n+1}}{\tau_{\alpha,n+1}} (\fsigma_{n+1}:D_\omega \Ms) \right) 
                       }_{=:D_\omega \gammaacc_{n+1}} \fomega_\Delta.
\end{align}
}
Next, extending the linearization of the residual (compare Eq.~\eqref{deltaR}) yields
\begin{equation}
 \Delta \fR = \pd{\fR}{\fS_{n+1}} \Delta \fS_{n+1} + \pd{\fR}{\fZ}\Delta \fZ + D_\omega\fR \, \fomega_\Delta =\fzero \ \ \ \Rightarrow 
 D_\omega \fS_{n+1} = -\left( \pd{\fR}{\fS_{n+1}} \right)^{-1} D_\omega\fR \nonumber
\end{equation}
with (see Eqns.~\eqref{rsigma} and \eqref{rtauc})
\begin{equation}
 D_\omega\fR = \begin{pmatrix} \ffC:D_\omega \fepsp_{n+1} \\ (\ttauc)' (\gammaacc_{n+1}) \, D_\omega \gammaacc_{n+1} \end{pmatrix}.
\end{equation}
Finally, the 'total' linearization of the internal variables with respect to the crystal orientation is described by (with a little abuse of notation)
\begin{equation}
 D_\omega \fX_{n+1} = \pd{\fX_{n+1}}{\fS_{n+1}} D_\omega \fS_{n+1} + \begin{pmatrix} D_\omega \fepsp_{n+1} \\ D_\omega \gammaacc_{n+1} \end{pmatrix}.
\end{equation}
\section{Modified latin hypercube sampling procedure}\label{applhc}
The strain deviator \f{\feps_0'} reads in Mandel notation
\begin{equation}
 \und{\hat \eps}'_0 = \sum\limits_{i=1}^5 \eps_i \und{\hat E}_i
\end{equation}
with orthonormal base vectors \f{\und{\hat E}_1=\sqrt{2/3}\T{(1,-0.5,-0.5,0,0,0)}}, \f{\und{\hat E}_2=\sqrt{2/2}\T{(0,-1,1,0,0,0,0)}}, \f{\und{\hat E}_3=\T{(0,0,0,1,0,0)}}, \f{\und{\hat E}_4=\T{(0,0,0,0,1,0)}} and \f{\und{\hat E}_5=\T{(0,0,0,0,0,1)}}. In a first step, the components \f{\eps_i} are taken from a conventional five-dimensional latin hypercube with hyperedge lengt 2. Subsequentlyl, all samples outside of a corresponding hypersphere (hyperradius 1) are thrown away. A small random volumetric contribution \f{\eps^\circ \fI} (\f{\fI} is the identity tensor) is added to each sample, where \f{\eps^\circ\in[-0.005,0.005]} is an equally distributed random number. Finally, the samples are 1normalized, such that \f{\|\feps_0\|=0.05}.
\end{appendix}
\bibliographystyle{elsarticle-harv}
\bibliography{lit}

\end{document}